\documentclass[11pt]{article}
\usepackage[square]{natbib}
\usepackage{graphicx}
\usepackage{dcolumn}
\usepackage{bm}
\usepackage{amsmath}
\usepackage{amsxtra}
\usepackage{amstext}
\usepackage{amssymb}
\usepackage{latexsym}

\begin{document}

\title{
\vskip-2.5truecm
\rightline{\small{\tt ULB-TH/08-15}}
\vskip2.5truecm
{ Thermodynamic Field Theory\\
with the Iso-Entropic Formalism}
}

\author{ Giorgio SONNINO}
\date{}
\maketitle
\begin{center}
{\it 
Department of Statistical Physics and Plasma\\
EURATOM - Belgian State Fusion Association\\
Universit{\'e} Libre de Bruxelles (U.L.B.)\\
Bvd du Triomphe, Campus de la Plaine, C.P. 231, Building  NO\\
Brussels, B-1050, Belgium\\
}
E-mail: {\tt gsonnino@ulb.ac.be}
\end{center}

\begin{abstract}
A new formulation of the thermodynamic field theory (TFT) is presented. In this new version, one of the basic restriction in the old theory, namely a closed-form solution for the thermodynamic field strength, has been removed. In addition, the general covariance principle is replaced by Prigogine's thermodynamic covariance principle (TCP). The introduction of TCP required the application of an appropriate mathematical formalism, which has been referred to as the {\it iso-entropic formalism}. The validity of the Glansdorff-Prigogine Universal Criterion of Evolution,  via geometrical arguments, is proven. A new set of thermodynamic field equations, able to determine the nonlinear corrections to the linear ("Onsager") transport coefficients, is also derived. The geometry of the thermodynamic space is non-Riemannian tending to be Riemannian for hight values of the entropy production.  In this limit, we obtain again the same thermodynamic field equations found by the old theory. Applications of the theory, such as transport in magnetically confined plasmas, materials submitted to temperature and electric potential gradients or to unimolecular triangular chemical reactions can be found at references cited herein.

\noindent 
\end{abstract}


\section{Introduction}

It is known that, close to equilibrium, the transport equations (i.e. the flux-forces equations) of a thermodynamic system are provided by the Onsager relations. Indicating with $X^\mu$ and $J_\mu$ the thermodynamic forces and fluxes, respectively, the Onsager relations read
\begin{equation}\label{i1}
J_\mu=\tau_{0\mu\nu}X^\nu
\end{equation}
\noindent where $\tau_{0\mu\nu}$ are the transport coefficients. We suppose that all quantities involved in Eqs~(\ref{i1}) are written in dimensionless form. Matrix $\tau_{0\mu\nu}$ can be decomposed into a sum of two matrices, one symmetric and the other skew-symmetric, which we denote with $L_{\mu\nu}$ and $f_0{\mu\nu}$, respectively. The second principle of thermodynamics imposes that $L_{\mu\nu}$ be a positive definite matrix. In this equation, as in the remainder of this paper, the Einstein summation convention on the repeated indexes is adopted. The most important property of Eqs~(\ref{i1}) is that near equilibrium, the coefficients $\tau_{\mu\nu}$ are independent of the thermodynamic forces, so that
\begin{equation}\label{i2}
\frac{\partial\tau_{0\mu\nu}}{\partial X^\lambda}=0
\end{equation}
\noindent The region where Eqs~(\ref{i2}) hold, is called {\it Onsager's region} or, the {\it linear region}. Many important theorems have been demonstrated for thermodynamic systems in the Onsager region. Among them, the most important one is the {\it Minimum Entropy Production Theorem}, showed by Prigogine in 1945-1947 \cite{prigogine}. This theorem establishes that, in the Onsager region, for $a-a$ processes\footnote{Here, we adopt the De Groot-Mazur terminology \cite{degroot}: when the velocity's distribution function of particles is an even (odd) function of the velocities of particles, the processes is referred to as a {\it $a$-process} ({\it $b$-process}). It is possible to show that this definition implies that the $a$-processes only involve the symmetric part of the Onsager matrix whereas the $b$-processes only the skew-symmetric one. \label{process}}, a thermodynamic system relaxes towards a steady-state in such a way that the rate of the entropy production is negative
\begin{equation}\label{i3}
\frac{d\sigma}{dt}\leq0\qquad \Bigl(\frac{d\sigma}{dt}=0\ \ {\rm at\ the\ steady\ state}\Bigr)
\end{equation}
\noindent where $\sigma=L_{\mu\nu}X^\mu X^\nu$ indicates the entropy production and $t$ is time. In 1954, Glansdorff and Prigogine demonstrated a more general theorem, valid also when the system is out of the Onsager region \cite{prigogine1}. They showed that, regardless of the type of processes, a thermodynamic system relaxes towards a steady-state in such a way that the following quantity $\mathcal{P}$ is negative 
\begin{equation}\label{i4}
\mathcal{P}\equiv J_\mu\frac{d X^\mu}{dt}\leq0\qquad \Bigl(\mathcal{P}=0\ \ {\rm at\ the\ steady\ state}\Bigr)
\end{equation}
\noindent Inequality (\ref{i4}) reduces to inequality (\ref{i3}) for $a-a$ processes in the Onsager region. For spatially-extended systems, the expression in Eqs.~(\ref{i4}) should be replaced by 
\begin{equation}\label{i5}
\mathcal{P}\equiv\int_\Omega {\mathcal J}_\mu\frac{d {\mathcal X}^\mu}{dt}dv\leq0\qquad \Bigl(\mathcal{P}=0\ \ {\rm at\ the\ steady\ state}\Bigr)
\end{equation}
\noindent where $dv$ is the volume element and the integration is over the entire space $\Omega$ occupied by the system in question. ${\mathcal J}_\mu({\bf r},t)$ and ${\mathcal X}^\mu({\bf r},t)$ denote the space-time dependent fluxes and forces, respectively. The inequality expressed in (\ref{i4}) [or in (\ref{i5})] is referred to as the {\it Universal Criterion of Evolution} and it is the most general result obtained up to now in thermodynamics of irreversible processes. Out of the Onsager region, the transport coefficients may depend on the thermodynamic forces and Eqs~(\ref{i2}) may loose their validity. Transport, in the nonlinear region, has been largely studied, both experimentally and theoretically. In particular, many theories, based on the Fourier expansion of the transport coefficients in terms of the thermodynamic forces, have been proposed (see, for example, refs \cite{gyarmati}, \cite{li} and \cite{rysselberghe}). The theoretical predictions are however in disagreement with the experiments and this is mainly due to the fact that, in the series expansion, the terms of superior order are greater than those of inferior order. Therefore truncation of the series at some order is not mathematically justified. 

\noindent A thermodynamic field theory (TFT) has been proposed in 1999 in order to evaluate how the relations between fluxes and forces, Eqs~(\ref{i1}), {\it deform} when the thermodynamic system is far from the linear ("Onsager") region \cite{sonnino}. Attempts to derive a generally covariant thermodynamic field theory (GTFT) can be found in refs \cite{sonnino1}. The characteristic feature of the TFT is its purely macroscopic nature. This does not mean a formulation based on the macroscopic evolution equations, but rather a purely thermodynamic formulation starting solely from the entropy production and from the transport equations, i.e., the fluxes-forces relations. The latter provide the possibility of defining an abstract space (the thermodynamic space), covered by the $n$ independent thermodynamic forces $X^\mu$, whose metric is identified with the symmetric part of the transport matrix. The law of evolution is not the dynamical law of particle motion, or the set of two-fluid macroscopic equations of plasma dynamics. The evolution in the thermodynamic space is rather determined by postulating three purely geometrical principles: the {\it Shortest Path Principle}, the {\it Thermodynamic Field Strength in closed form}, and the {\it Principle of Least Action}. From theses principles, a set of field equations, constraints, and boundary conditions are derived. These equations, referred to as the {\it thermodynamic field equations}, determine the nonlinear corrections to the linear ("Onsager") transport coefficients. However, the formulation of the thermodynamic field theory, as reported in refs~\cite{sonnino}, raises the following fundamental objection
\vskip0.2truecm
\noindent {\it There are no strong experimental evidences supporting the requirement that the thermodynamic field strength is in a closed form}.
\vskip0.2truecm
\noindent Moreover, the principle of general covariance, which in refs \cite{sonnino1} has been  assumed to be valid for general transformations in the space of thermodynamic configurations, is, in reality, respected only by a very limited class of thermodynamic processes. In this paper, through an appropriate mathematical formalism, the {\it iso-entropic formalism}, the entire TFT is re-formulated removing the assumptions regarding the closed-form of the thermodynamic field strength and the general covariance principle (GCP). The GCP is replaced by  the {\it thermodynamic covariance principle}  (TCP), or the Prigogine statement \cite{prigogine}, establishing that thermodynamic systems, obtained by a transformation of forces and fluxes in such a way that the entropy production remains unaltered, are thermodynamically equivalent. This principle applies to transformations in the thermodynamic space, and they are referred to as the {\it  thermodynamic coordinate transformations} (TCT). It is worthwhile mentioning that the TCP is actually largely used in a wide variety of thermodynamic processes ranging from non equilibrium chemical reactions to transport processes in tokamak plasmas (see, for examples, the papers and books cited in refs \cite{hinton}) and \cite{balescu2}). To the author knowledge, the validity of the thermodynamic covariance principle has been verified empirically without exception in physics until now.  

\noindent The analysis starts from the following observation. Consider a relaxation process of a thermodynamic system in the Onsager region. If the system relaxes towards a steady-state along the shortest path in the thermodynamic space, then the Universal Criterion of Evolution is automatically satisfied. Indeed, in this case, we can write 
\begin{equation}\label{i6}
J_\mu{\dot X}^\mu=(L_{\mu\nu}+f_{0\mu\nu})X^\nu{\dot X^\mu}
\end{equation}
\noindent where the dot over the variables indicates the derivative with respect to the arc parameter $\varsigma$, defined as 
\begin{equation}\label{i7}
d\varsigma^2=(L_{\mu\nu}dX^\mu dX^\nu)^{1/2}
\end{equation}
\noindent Parameter $\varsigma$ can be chosen in such a way that it vanishes when the system begins to evolve and it assumes the value, say $l$, when the system reaches the steady-state. In the Onsager region, the thermodynamic space is an Euclidean space with metric $L_{\mu\nu}$. The equation of the shortest path reads ${\ddot X}^\mu=0$, with solution of the form
\begin{equation}\label{i8}
X^\mu=a^\mu\varsigma +b^\mu
\end{equation}
\noindent where $a^\mu$ and $b^\mu$ are arbitrary constant independent of the arc parameter. Inserting Eq.~(\ref{i8}) into Eq.~(\ref{i6}) and observing that  $L_{\mu\nu}a^\mu a^\nu=1$ and $f_{0\mu\nu}a^\mu a^\nu=0$, we find
\begin{equation}\label{i9}
J_\mu{\dot X}^\mu=\varsigma+\tau_{0\mu\nu}a^\mu b^\nu
\end{equation}
\noindent At the steady state (i.e. for $\varsigma=l$) $J_\mu{\dot X}^\mu\mid_{st.state}=0$ (because $\mathcal{P}\mid_{st.state}=0$). Eq.~(\ref{i9}) can then be written as
\begin{equation}\label{i10}
P=-(l-\varsigma)\leq0\qquad({\rm with}\quad P\equiv J_{\mu}{\dot X}^\mu)
\end{equation}
\noindent or
\begin{equation}\label{i11}
\mathcal{P}=-(l-\varsigma)\Bigl(L_{\mu\nu}\frac{dX^\mu}{dt}\frac{dX^\nu}{dt}\Bigr)^{1/2}\leq0
\end{equation}
\noindent The equation for the dissipative quantity $P$, when the thermodynamic system relaxes in the linear region, is thus given by Eq.~(\ref{i9}):
\begin{equation}\label{i12}
\frac{dP}{d\varsigma}=1
\end{equation}
\noindent Also note that ${\dot\sigma}=2P\leq0$ i.e., the minimum entropy production theorem is also satisfied during relaxation. Outside of Onsager's region, one may be tempted to construct a Riemannian space (of $3$ or more dimensions) which is projectively flat i.e., having a vanishing Weyl's projective curvature tensor. In this case, indeed, there exists a coordinate system such that the equations of the shortest path are linear in the coordinates [i.e., the shortest paths are given by equations of the form (\ref{i8})]. In this respect, we have the following Weyl theorem \cite{weyl}: a necessary and sufficient condition that a Riemannian space be projectively flat is that its Riemannian curvature be constant everywhere. On the other hand, to re-obtain the Onsager relations, we should also require that, near equilibrium, the Riemannian space reduces to a flat space (which has zero Riemannian curvature). The Weyl theorem can be conciliated with our request only if there exists a coordinate system such that Eqs~(\ref{i2}) are valid everywhere, which is in contrast with experiments. Thus one wants the Universal Criterion of Evolution satisfied also out of the Onsager region, without imposing a priori {\it any} restrictions on transport coefficients, a {\it non-Riemannian} thermodynamic space is required. Clearly, a transport theory without a knowledge of microscopic dynamical laws can not be developed. Transport theory is only but an aspect of non-equilibrium statistical mechanics, which provides the link between micro and macro-levels. This link appears indirectly in the "unperturbed" matrices, i.e. the $L_{\mu\nu}$ and the $f_{0\mu\nu}$ coefficients, used as an input in the equations: these coefficients have to be calculated in the usual way by kinetic theory.

\noindent In section \ref{isoef}, we introduce a non-Riemannian space whose geometry is constructed in such a way that
\begin{description}
\item {\it {\bf A.} The theorems valid when a generic thermodynamic system relaxes out of equilibrium are satisfied};
\item {\it {\bf B.} The differential equations for the transport coefficients are covariant under the thermodynamic coordinate transformations (TCT)}.
\end{description}
\noindent We shall see that the properties of geometry do not depend on the shortest paths but upon {\it a particular expression of the affine connection}. Our geometry is then of {\it affine type} and not of projective type. At the end of section \ref{isoef} we derive the field equations for the transport coefficients through an appropriate mathematical formalism: the {\it iso-entropic formalism}. This formalism allows to respect the Prigogine statement. New objects like {\it thermodynamic covariant differentiation} or the {\it thermodynamic curvature} are also introduced. We shall see that under the weak-field approximation and when $\sigma\gg 1$, but only in these limits, the new thermodynamic field equations reduce to the ones obtained in refs \cite{sonnino}. So that, all results found in refs \cite{sonnino3}, for magnetically confined plasmas, and in refs \cite{sonnino4}, for the nonlinear thermoelectric effect and the unimolecular triangular reaction, remain valid. In section \ref{ttse} it is shown that this formalism is able to verify the thermodynamic theorems (in particular, the Universal Criterion of Evolution) for systems relaxing out of the Onsager region. Mathematical details and demonstrations of the theorems are reported in the annexes.

\vskip 0.2truecm
\section{The Iso-Entropic Formalism}\label{isoef}
\vskip 0.2truecm
\noindent Consider a thermodynamic system driven out from equilibrium by a set of $n$ independent thermodynamic forces $\{X^\mu\}$ ($\mu=1,\cdots n$). It is also assumed that the system is submitted to time-independent boundary conditions. The set of conjugate flows, $\{J_{\mu}\}$, is coupled to the thermodynamic forces through the relation 
\begin{equation}\label{ief0}
J_{\mu}=\tau_{\mu\nu}X^\nu
\end{equation}
\noindent where $\tau_{\mu\nu}$ denote the transport coefficients. The symmetric piece of $\tau_{\mu\nu}$ is denoted with $g_{\mu\nu}$ and the skew-symmetric piece as $f_{\mu\nu}$:
\begin{equation}\label{ief1}
\tau_{\mu\nu}=\frac{1}{2}(\tau_{\mu\nu}+\tau_{\nu\mu})+\frac{1}{2}(\tau_{\mu\nu}-\tau_{\nu\mu})=g_{\mu\nu}+f_{\mu\nu}
\end{equation}
\noindent where
\begin{eqnarray}\label{ief2}
&&g_{\mu\nu}=\frac{1}{2}(\tau_{\mu\nu}+\tau_{\nu\mu})=g_{\nu\mu}\\
&&f_{\mu\nu}=\frac{1}{2}(\tau_{\mu\nu}-\tau_{\nu\mu})=-f_{\nu\mu}
\end{eqnarray}
\noindent  It is assumed that $g_{\mu\nu}$ is a positive definite matrix. With the elements of the transport coefficients two objects are constructed: {\it operators}, which may act on thermodynamic tensorial objects and {\it thermodynamic tensorial objects}, which under coordinate (forces) transformations, obey to well specified transformation rules.
\vskip 0.2truecm
\noindent{\bf Operators}
\vskip 0.2truecm
Two operators are introduced, the {\it entropy production operator} $\sigma (X)$ and the {\it dissipative quantity operator} $P(X)$, acting on the thermodynamic forces in the following manner
\begin{eqnarray}\label{o1}
&&\sigma (X):\rightarrow\sigma(X)\equiv X gX^T\nonumber\\
&& P(X):\rightarrow P(X)\equiv X\tau {\dot X}^T
\end{eqnarray}
\noindent In Eqs~(\ref{o1}), the transport coefficients are then considered as elements of the two $n$ x $n$ matrices, $\tau$ and $g$. The positive definiteness of the matrix $g_{\mu\nu}$ ensures the validity of the second principle of thermodynamics: $\sigma\ge0$. These matrices multiply the thermodynamic forces $X$ expressed as $n$ x $1$ column matrices. The dot symbol stands for derivative with respect to parameter $\varsigma$, defined in Eq.~(\ref{tct4}). Thermodynamic states $X_{s}$ such that
\begin{eqnarray}\label{o2}
P(X_{s})=0
\end{eqnarray}
\noindent are referred to as {\it steady-states}. These are physical quantities and should remain invariant under thermodynamic coordinate transformations. Eqs~(\ref{o1}) {\it should not} be interpreted as the metric tensor $g_{\mu\nu}$, which acts on the coordinates. The metric tensor {\it acts only on} elements of the tangent space (like $dX^\mu$, see the forthcoming paragraphs) or on the thermodynamic tensorial objects.
\vskip 0.2truecm
\noindent{\bf Transformation Rules of Entropy Production, Forces and Flows}
\vskip 0.2truecm
According to Prigogine's statement \cite{prigogine}, {\it thermodynamic systems are thermodynamically equivalent if, under transformation of fluxes and forces the bilinear form of the entropy production, $\sigma$, remains unaltered}. In mathematical terms, this implies:
 \begin{equation}\label{tr1}
 \sigma=J_\mu X^\mu=J'_\mu X'^\mu
 \end{equation}
 \noindent This condition requires that the transformed thermodynamic forces and flows satisfy the relation
\begin{eqnarray}\label{tr2}
&&X'^\mu=\frac{\partial X'^{\mu}}{\partial X^\nu} X^\nu\nonumber\\
&& J'_\mu=\frac{\partial X^{\nu}}{\partial X'^\mu}J_\nu
\end{eqnarray}
\noindent These transformations are referred to as {\it Thermodynamic Coordinate Transformations} (TCT). The expression of entropy production becomes accordingly
 \begin{equation}\label{tr3}
 \sigma=J_\mu X^\mu=\tau_{\mu\nu}X^\mu X^\nu=g_{\mu\nu}X^\mu X^\nu=g'_{\mu\nu}X'^\mu X'^\nu=\sigma'
 \end{equation}
 \noindent From Eq.~(\ref{tr3}) we find
 \begin{equation}\label{tr4}
g'_{\lambda\kappa}=g_{\mu\nu}\frac{\partial X^\mu}{\partial X'^\lambda}
 \frac{\partial X^\nu}{\partial X'^\kappa}
\end{equation}
\noindent Moreover, inserting Eqs~(\ref{tr2}) and Eq.~(\ref{tr4}) into relation $J_{\mu}=(g_{\mu\nu}+f_{\mu\nu})X^\nu$, we obtain
\begin{equation}\label{tr5}
J'_\lambda=\Bigl(g'_{\lambda\kappa}+f_{\mu\nu}\frac{\partial X^\mu}{\partial X'^\lambda}\frac{\partial X^\nu}{\partial X'^\kappa}\Bigr)X'^\kappa
\end{equation}
\noindent or 
 \begin{equation}\label{tr6}
J'_\lambda=(g'_{\lambda\kappa}+f'_{\lambda\kappa})X'^\kappa\quad\qquad{\rm with}\qquad
f'_{\lambda\kappa}=f_{\mu\nu}\frac{\partial X^\mu}{\partial X'^\lambda}
 \frac{\partial X^\nu}{\partial X'^\kappa}
\end{equation}
\noindent Hence, the transport coefficients transform like a {\it thermodynamic tensor of second order}
\footnote{We may qualify as {\it thermodynamic tensor} (taken as a single noun) a set of quantities where {\it only transformations Eqs~(\ref{tr2}) are involved}. This is in order to qualify as a {\it tensor}, a set of quantities, which satisfies certain laws of transformation when the coordinates undergo a general transformation. Consequently every tensor is a thermodynamic tensor but the converse is {\it not} true.\label{thermotensor}}.

\vskip0.5truecm
\noindent{\bf Properties of the TCT}
\vskip0.5truecm

\noindent By direct inspection, it is easy to verify that the general solutions of equations (\ref{tr2}) are 
\begin{equation}\label{tct1}
X'^\mu=X^1F^\mu\Bigl(\frac{X^2}{X^1},\ \frac{X^3}{X^2},\ \cdots\ \frac{X^n}{X^{n-1}}\Bigr)
\end{equation}
\noindent where $F^\mu$ is an {\it arbitrary function} of variables $X^j/X^{j-1}$ with ($j=2,\dots, n$). We may (or we must) require that, in the Onsager region, Eqs~(\ref{tct1}) reduce to linear homogeneous transformations
\begin{equation}\label{tct1a}
X'^\mu=c_\nu^\mu X^\nu
\end{equation}
\noindent where $c_\nu^\mu$ are constant coefficients (i.e., independent of the thermodynamic forces). Note that from Eq.(\ref{tr2}), the following important identities are derived
\begin{equation}\label{tct2}
X^\nu\frac{\partial^2X'^\mu}{\partial X^\nu\partial X^\kappa}=0\qquad ;\qquad X'^\nu\frac{\partial^2X^\mu}{\partial X'^\nu\partial X'^\kappa}=0
\end{equation}
\noindent Moreover
\begin{eqnarray}\label{tct3}
&&dX'^\mu=\frac{\partial X'^{\mu}}{\partial X^\nu} dX^\nu \nonumber\\
&&\frac{\partial}{\partial X'^\mu}= \frac{\partial X^{\nu}}{\partial X'^\mu}\frac{\partial}{\partial X^\nu} 
\end{eqnarray}
\noindent i.e., $dX^\mu$ and $\partial/\partial X^{\mu}$ transform like a thermodynamic contra-variant and a thermodynamic covariant vector, respectively. According to Eq.~(\ref{tct3}), thermodynamic vectors $dX^\mu$ define the {\it tangent space} to $Ts$. It also follows that the operator $P(X)$, i.e. the dissipation quantity, and in particular the definition of steady-states, are invariant under TCT. Parameter $\varsigma$, defined as 
\begin{equation}\label{tct4}
d\varsigma^2=g_{\mu\nu}dX^\mu dX^\nu
\end{equation}
\noindent is a scalar under TCT. The operator $\mathcal{O}$
\begin{equation}\label{pts14d}
\mathcal{O}\equiv X^\mu\frac{\partial}{\partial X^\mu}=X'^\mu\frac{\partial}{\partial X'^\mu}=\mathcal{O}'
\end{equation}
\noindent is also invariant under TCT. This operator plays an important role in the formalism. 
\vskip 0.2truecm
\noindent{\bf Thermodynamic Space, Thermodynamic Covariant Derivatives and Thermodynamic Curvature}
\vskip 0.2truecm
A non-Riemannian space with a linear connection $\Gamma_{\alpha\beta}^\mu$ is now introduced. Consider an $n$-space in which the set of quantities $\Gamma_{\alpha\beta}^\mu$ is assigned as functions of the $n$ independent thermodynamic forces $X^\mu$, chosen as coordinate system. Under a coordinate (forces) transformation, it is required that the functions $\Gamma_{\alpha\beta}^\mu$ transform according to the law
\begin{equation}\label{ts1}
{\Gamma'}_{\alpha\beta}^\mu=\Gamma_{\lambda\kappa}^\nu\frac{\partial X'^\mu}{\partial X^\nu}\frac{\partial X^\lambda}{\partial X'^\alpha}\frac{\partial X^\kappa}{\partial X'^\beta}+\frac{\partial X'^\mu}{\partial X^\nu}\frac{\partial^2X^\nu}{\partial X'^\alpha \partial X'^\beta}
\end{equation}
\noindent With the linear connection $\Gamma_{\alpha\beta}^\mu$, the absolute derivative of a thermodynamic contra-variant vector $T^\mu$ along a curve can be defined as
\begin{equation}\label{ts2}
\frac{\delta T^\mu}{\delta\varsigma}=\frac{dT^\mu}{d\varsigma}+\Gamma^\mu_{\alpha\beta}T^\alpha\frac{dX^\beta}{d\varsigma}
\end{equation}
\noindent It is easily checked that, if the parameter along the curve is changed from $\varsigma$ to $\varrho$, then the absolute derivative of a thermodynamic tensor field with respect to $\varrho$ is $d\varsigma/d\varrho$ times the absolute derivative with respect to $\varsigma$. The absolute derivative of any contra-variant thermodynamic tensor may be easily obtained generalizing Eq.~(\ref{ts2}). In addition, the linear connection $\Gamma^\mu_{\alpha\beta}$ is submitted to the following basic postulates:
\begin{description}
\item{\it {\bf 1.} The absolute derivative of a thermodynamic contra-variant tensor is a thermodynamic tensor of the same order and type.}
\item{\it {\bf 2.} The absolute derivative of an outer product of thermodynamic tensors, is given, in terms of factors, by the usual rule for differentiating a product.}
\item{\it {\bf 3.} The absolute derivative of the sum of thermodynamic tensors of the same type is equal to the sum of the absolute derivatives of the thermodynamic tensors.}
\end{description}
\noindent In a space with a linear connection, we can introduce the notion of the {\it shortest path} defined as {\it a curve such that a thermodynamic vector, initially tangent to the curve and propagated parallelly along it, remains tangent to the curve at all points}. By a suitable choice of the parameter $\varrho$, the differential equation for the shortest path is simplified reducing to
\begin{equation}\label{ts3}
\frac{d^2X^\mu}{d\varrho^2}+\Gamma^\mu_{\alpha\beta}\frac{dX^\alpha}{d\varrho}\frac{dX^\beta}{d\varrho}=0
\end{equation}
\noindent To respect the general requirement ${\bf A.}$ (see section \ref{ttse}), it is required that the absolute derivative of the entropy production satisfies the equality 
\begin{equation}\label{ts4}
\frac{\delta\sigma}{\delta\varsigma}=J_\mu\frac{\delta X^\mu}{\delta\varsigma}+X^\mu\frac{\delta J_\mu}{\delta\varsigma}
\end{equation}
\noindent More in general, it is required that the operations of contraction and absolute differentiation commute for all thermodynamic vectors. As a consequence, the considered space should be a space with a single connection. The absolute derivative of a covariant thermodynamic vector $T_\mu$ is then defined as 
\begin{equation}\label{ts5}
\frac{\delta T_\mu}{\delta\varsigma}=\frac{dT_\mu}{d\varsigma}-\Gamma^\alpha_{\mu\beta}T_\alpha\frac{dX^\beta}{d\varsigma}
\end{equation}
\noindent The absolute derivative of the most general contra-variant, covariant and mixed thermodynamic tensors may be obtained generalizing Eqs~(\ref{ts2}) and (\ref{ts5}). 

\noindent The derivatives, covariant under TCT, of thermodynamic vectors, are defined as
\begin{eqnarray}\label{ts6}
&&T^\mu_{\mid\nu}=\frac{\partial T^\mu}{\partial X^\nu}+\Gamma^\mu_{\alpha\nu}T^\alpha\nonumber\\
&&T_{\mu\mid\nu}=\frac{\partial T_\mu}{\partial X^\nu}-\Gamma^\alpha_{\mu\nu}T_\alpha
\end{eqnarray}
\noindent For the entropy production, it is also required that
\begin{equation}\label{ts7}
\sigma_{\mid\mu\mid\nu}=\sigma_{\mid\nu\mid\mu}
\end{equation}
\noindent More in general, Eq.~(\ref{ts7}) should be verified for any thermodynamic scalar $T$. This postulate requires that the linear single connection $\Gamma^\mu_{\alpha\beta}$ is also symmetric i.e., $\Gamma^\mu_{\alpha\beta}=\Gamma^\mu_{\beta\alpha}$. A non-Riemannian geometry can now be constructed out of $n^2(n+1)/2$ quantities, the components of $\Gamma^\mu_{\alpha\beta}$, according to the general requirements  ${\bf A}$ and ${\bf B}$ mentioned in the introduction.

\noindent In the forthcoming paragraph, the expression of the affine connection $\Gamma^\mu_{\alpha\beta}$ is determined from assumption ${\bf A}$. In section  \ref{ttse} it is shown that the Universal Criterion of Evolution, applied to thermodynamic systems relaxing towards a steady-state, is automatically satisfied along the shortest path if, in case of symmetric processes (i.e., for $a-a$ processes), we impose
\begin{equation}\label{ts8}
\Gamma^\mu_{\alpha\beta}=\frac{1}{2}g^{\mu\lambda}\Bigl(\frac{\partial g_{\lambda\alpha}}{\partial X^\beta}+\frac{\partial g_{\lambda\beta}}{\partial X^\alpha}-\frac{\partial g_{\alpha\beta}}{\partial X^\lambda}\Bigr)+\frac{1}{2\sigma}X^\mu\mathcal{O}(g_{\alpha\beta})
\end{equation}
\noindent In the general case, we have 
\begin{eqnarray}\label{ts9}
\!\!\!\!\!\!\!\!\!\!\!\!\!\!\!\Gamma^\mu_{\alpha\beta}=\!\!\!\!\!\!\!\!\!&&{\bar N}^{\mu\kappa}g_{\kappa\lambda}\begin{Bmatrix} 
\lambda \\ \alpha\beta
\end{Bmatrix}+\frac{{\bar N}^{\mu\kappa}}{2\sigma}X_\kappa\mathcal{O}(g_{\alpha\beta})+\frac{{\bar N}^{\mu\kappa} }{2\sigma}X_\kappa X^\lambda\Bigl(\frac{\partial f_{\alpha\lambda}}{\partial X^\beta}+\frac{\partial f_{\beta\lambda}}{\partial X^\alpha}\Bigr)\nonumber\\
&&
\qquad\qquad\qquad\ \ \!
+\frac{{\bar N}^{\mu\kappa}}{2\sigma}f_{\kappa\varsigma}X^\varsigma X^\lambda\Bigl(\frac{\partial g_{\alpha\lambda}}{\partial X^\beta}+\frac{\partial g_{\beta\lambda}}{\partial X^\alpha}\Bigr)
\end{eqnarray}
\noindent where the thermodynamic Christoffel symbols of the second kind are introduced
\begin{equation}\label{ts10}
\begin{Bmatrix} 
\mu \\ \alpha\beta
\end{Bmatrix}=\frac{1}{2}g^{\mu\lambda}\Bigl(\frac{\partial g_{\lambda\alpha}}{\partial X^\beta}+\frac{\partial g_{\lambda\beta}}{\partial X^\alpha}-\frac{\partial g_{\alpha\beta}}{\partial X^\lambda}\Bigr)
\end{equation}
\noindent and matrix ${\bar N}^{\mu\kappa}$ is defined as
\begin{equation}\label{ts11}
\!\ \!N_{\mu\nu}\equiv g_{\mu\nu}+\frac{1}{\sigma}f_{\mu\kappa}X^\kappa X_\nu+\frac{1}{\sigma}f_{\nu\kappa}X^\kappa X_\mu\quad{\rm with}\ \ \ {\bar N^{\mu\kappa}}:\ \ {\bar N}^{\mu\kappa}N_{\nu\kappa}=\delta_{\nu}^{\mu}
\end{equation}
\noindent In appendix \ref{appx} it is proven that the affine connections Eqs~(\ref{ts8}) and (\ref{ts9}) transform, under a TCT, as in Eq.~(\ref{ts1}) and satisfy the postulates ${\bf 1.}$, ${\bf 2.}$ and ${\bf 3.}$ From Eq.~(\ref{ts11}) we obtain 
\begin{eqnarray}\label{ts12}
\!\!\!\!\!\!\!\!\!\!\!\!\!\!\!\!\!\!\!\!\!\!\!\!\!\!\!\!&&N_{\mu\nu}=N_{\nu\mu}\nonumber\\
\!\!\!\!\!\!\!\!\!\!\!\!\!\!\!\!\!\!\!\!\!\!\!\!\!\!\!\!&&N_{\mu\nu}X^\nu=\Bigl(g_{\mu\nu}+\frac{1}{\sigma}f_{\mu\kappa}X^\kappa X_\nu+\frac{1}{\sigma}f_{\nu\kappa}X^\kappa X_\mu \Bigr)X^\nu
=J_{\mu}\nonumber\\
\!\!\!\!\!\!\!\!\!\!\!\!\!\!\!\!\!\!\!\!\!\!\!\!\!\!\!\!&&N_{\mu\nu}X^\mu=\Bigl(g_{\mu\nu}+\frac{1}{\sigma}f_{\mu\kappa}X^\kappa X_\nu+\frac{1}{\sigma}f_{\nu\kappa}X^\kappa X_\mu \Bigr)X^\mu=J_\nu\\
\!\!\!\!\!\!\!\!\!\!\!\!\!\!\!\!\!\!\!\!\!\!\!\!\!\!\!\!&&N_{\mu\nu} X^\nu X^\mu
=J_\mu X^\mu=\sigma\nonumber
\end{eqnarray}
\noindent While
\begin{eqnarray}\label{ts13}
\!\!\!\!\!\!\!\!\!\!\!\!\!\!\!\!\!\!\!\!\!\!\!\!\!\!\!\!\!\!\!\!\!\!\!\!\!\!\!\!\!\!\!\!\!\!\!\!\!\!\!\!\!\!\!\!\!\!\!
&&{\bar N}^{\mu\kappa}={\bar N}^{\kappa\mu}\nonumber\\
\!\!\!\!\!\!\!\!\!\!\!\!\!\!\!\!\!\!\!\!\!\!\!\!\!\!\!\!\!\!\!\!\!\!\!\!\!\!\!\!\!\!\!\!\!\!\!\!\!\!\!\!\!\!\!\!\!\!\!
&&{\bar N}^{\mu\kappa}J_{\mu}={\bar N}^{\mu\kappa}N_{\mu\nu}X^\nu={\bar N}^{\kappa\mu}N_{\nu\mu}X^\nu=X^\kappa\\
\!\!\!\!\!\!\!\!\!\!\!\!\!\!\!\!\!\!\!\!\!\!\!\!\!\!\!\!\!\!\!\!\!\!\!\!\!\!\!\!\!\!\!\!\!\!\!\!\!\!\!\!\!\!\!\!\!\!\!
&&{\bar N}^{\mu\kappa}J_{\kappa}={\bar N}^{\kappa\mu}J_{\kappa}=X^\mu\nonumber\\
\!\!\!\!\!\!\!\!\!\!\!\!\!\!\!\!\!\!\!\!\!\!\!\!\!\!\!\!\!\!\!\!\!\!\!\!\!\!\!\!\!\!\!\!\!\!\!\!\!\!\!\!\!\!\!\!\!\!\!
&&{\bar N}^{\mu\kappa}J_{\kappa}J_{\mu}=X^{\mu}J_{\mu}=\sigma\nonumber
\end{eqnarray}
\noindent The shortest path is the same for two symmetric connections whose coefficients are related as
\begin{equation}\label{ts14}
{\bar\Gamma}_{\alpha\beta}^\mu=\Gamma_{\alpha\beta}^\mu+\delta^\mu_\alpha\psi_\beta+\delta^\mu_\beta\psi_\alpha
\end{equation}
\noindent where $\psi_\alpha$ is an arbitrary covariant thermodynamic vector and $\delta^\mu_\nu$ denotes the Kronecker tensor. In literature, modifications of the connection similar to Eqs~(\ref{ts14}) are referred to as {\it projective transformations} of the connection and $\psi_\alpha$ the {\it projective covariant vector}. The introduction of the affine connection gives rise to the following difficulty:  the Universal Criterion of Evolution is satisfied for every shortest path constructed with affine connections ${\bar\Gamma}_{\alpha\beta}^\mu$, linked to $\Gamma_{\alpha\beta}^\mu$ by projective transformations. This leads to an indetermination of the expression for the affine connection, which is not possible to remove by using the Prigogine statement and the thermodynamic theorems alone. This problem can be solved by postulating that {\it the thermodynamic field equations} (i.e., the field equations for the affine connection and the transport coefficients) {\it be symmetric and projective-invariant} (i.e., invariant under projective transformations). 

\noindent For any covariant thermodynamic vector field $T_\mu$, we can form the thermodynamic tensor $R^{\mu}_{\nu\lambda\kappa}$ in the following manner \cite{synge}
\begin{equation}\label{ts15}
T_{\nu\mid\lambda\mid\kappa}-T_{\nu\mid\kappa\mid\lambda}=T_{\mu}R^{\mu}_{\nu\lambda\kappa}
\end{equation}
\vskip 0.2truecm
\noindent where 
\begin{equation}\label{ts16}
R^{\mu}_{\nu\lambda\kappa}=\frac{\partial \Gamma^\mu_{\nu\kappa}}{\partial X^\lambda}-\frac{\partial \Gamma^\mu_{\nu\lambda}}{\partial X^\kappa}+\Gamma^\eta_{\nu\kappa}\Gamma^\mu_{\eta\lambda}-\Gamma^\eta_{\nu\lambda}\Gamma^\mu_{\eta\kappa}
\end{equation}
\noindent  with $R^{\mu}_{\nu\lambda\kappa}$ satisfying the following identities
\begin{eqnarray}\label{ts17}
&&R^{\mu}_{\nu\lambda\kappa}=-R^{\mu}_{\nu\kappa\lambda}\nonumber\\
&&R^{\mu}_{\nu\lambda\kappa}+R^{\mu}_{\lambda\kappa\nu}+R^{\mu}_{\lambda\nu\kappa}=0\\
&&R^{\mu}_{\nu\lambda\kappa\mid\eta}+R^{\mu}_{\nu\kappa\eta\mid\lambda}+R^{\mu}_{\nu\eta\lambda\mid\kappa}=0\nonumber
\end{eqnarray}
\noindent By contraction, we obtain two distinct thermodynamic tensors of second order
\begin{eqnarray}\label{ts18}
&&R_{\nu\lambda}=R^{\mu}_{\nu\lambda\mu}=\frac{\partial \Gamma^\mu_{\nu\mu}}{\partial X^\lambda}-\frac{\partial \Gamma^\mu_{\nu\lambda}}{\partial X^\mu}+\Gamma^\eta_{\nu\mu}\Gamma^\mu_{\eta\lambda}-\Gamma^\eta_{\nu\lambda}\Gamma^\mu_{\eta\mu}\nonumber\\
&&F_{\lambda\nu}=\frac{1}{2}R^{\mu}_{\mu\lambda\nu}=\frac{1}{2}\Bigl(\frac{\partial \Gamma^\mu_{\nu\mu}}{\partial X^\lambda}-\frac{\partial \Gamma^\mu_{\lambda\mu}}{\partial X^\nu}\Bigr)
\end{eqnarray} with $F_{\lambda\nu}$ being skew-symmetric and $R_{\nu\lambda}$ asymmetric. Tensor $R_{\nu\lambda}$ can be re-written as 
\begin{eqnarray}\label{ts19}
&&\!\!\!\!\!\!\!\!\!\!\!\!\!\!\!\!R_{\nu\lambda}=B_{\nu\lambda}+F_{\lambda\nu}\qquad{\rm where}\nonumber\\
&&\!\!\!\!\!\!\!\!\!\!\!\!\!\!\!\!B_{\nu\lambda}=B_{\lambda\nu}=\frac{1}{2}\Bigl(\frac{\partial \Gamma^\mu_{\nu\mu}}{\partial X^\lambda}+
\frac{\partial \Gamma^\mu_{\lambda\mu}}{\partial X^\nu}\Bigr)
-\frac{\partial \Gamma^\mu_{\nu\lambda}}{\partial X^\mu}+\Gamma^\eta_{\nu\mu}\Gamma^\mu_{\eta\lambda}-\Gamma^\eta_{\nu\lambda}\Gamma^\mu_{\eta\mu}
\end{eqnarray}
\noindent Hence, $F_{\lambda\nu}$ is the skew-symmetric part of $R_{\nu\lambda}$. It is argued that the thermodynamic field equations can be derived by variation of a stationary action, which involves $R_{\nu\lambda}$. Symmetric and projective-invariant field equations may be obtained by proceeding in following manner: 1) a suitable projective transformation of the affine connection is derived so that $R_{\nu\lambda}$ be symmetric and $F_{\lambda\nu}$ be a zero thermodynamic tensor and 2) the most general projective transformation that leave unaltered $R_{\nu\lambda}$ and $F_{\lambda\nu}$ ($=0$) is determined. By a projective transformation, it is found that
\begin{eqnarray}\label{ts20}
&&{\bar B}_{\nu\lambda}=B_{\nu\lambda}+n\Bigl(\frac{\partial\psi_\nu}{\partial X^\lambda}-\psi_\nu\psi_\lambda\Bigr)-\Bigl(\frac{\partial\psi_\lambda}{\partial X^\nu}-\psi_\nu\psi_\lambda\Bigr)\nonumber\\
&&{\bar F}_{\lambda\nu}=F_{\lambda\nu}+\frac{n+1}{2}\Bigl(\frac{\partial\psi_\lambda}{\partial X^\nu}-\frac{\partial\psi_\nu}{\partial X^\lambda}\Bigr)
\end{eqnarray}
\noindent Eq.~(\ref{ts18}) shows that $F_{\lambda\nu}$ can be written as the curl of the vector $a_\nu/2$ defined as \cite{eisenhart} 
\begin{equation}\label{ts21}
a_\nu=\Gamma_{\kappa\nu}^\kappa-
\begin{Bmatrix} 
\kappa \\ \kappa\nu
\end{Bmatrix}
\end{equation}
\noindent Consequently, by choosing
\begin{equation}\label{ts22}
\psi_{\nu}=-\frac{1}{n+1}\Bigl(\Gamma_{\kappa\nu}^\kappa-
\begin{Bmatrix} 
\kappa \\ \kappa\nu
\end{Bmatrix}\Bigr)
\end{equation}
\noindent we have ${\bar F}_{\lambda\nu}=0$ and ${\bar R}_{\nu\lambda}={\bar B}_{\nu\lambda}$. From Eqs~(\ref{ts20}), we also have that the thermodynamic tensor ${\bar{\bar R}}_{\nu\lambda}$ remains symmetric for projective transformations of connection if, and only if, the projective covariant vector is the gradient of an arbitrary function of the $X$'s \cite{eisenhart}. In this case, the thermodynamic tensor ${\bar{\bar F}}_{\lambda\nu}$ remains unaltered i.e., ${\bar{\bar F}}_{\lambda\nu}=0$. Hence, at this stage, the expression of the affine connection is determined up to the gradient of a function, say $\phi$, of the thermodynamic forces, which is also {\it scalar under TCT}. Let us impose now the projective-invariance. Eqs~(\ref{ts20}) indicate that a necessary and sufficient condition that ${\bar R}_{\nu\lambda}$ be projective-invariant is that 
\begin{eqnarray}\label{ts23}
&&\!\!\!\!\!\!\!\!\!\!\!\!\!\!\!\!\!\!\!\!\!\!\!\!\!\!\!\!\!\!\!\!\!\!\!\!\!\!\!\!\!\!\!\!\!\!\!\!
\frac{\partial^2\phi}{\partial X^\lambda\partial X^\nu}-\frac{\partial\phi}{\partial X^\lambda}\frac{\partial\phi}{\partial X^\nu}=0\qquad{\rm with}\nonumber\\
&&\!\!\!\!\!\!\!\!\!\!\!\!\!\!\!\!\!\!\!\!\!\!\!\!\!\!\!\!\!\!\!\!\!\!\!\!\!\!\!\!\!\!\!\!\!\!\!\!
\left\{ \begin{array}{ll}
\phi=0 & \\
\frac{\partial\phi}{\partial X^\mu}=0 \\
\frac{\partial^2\phi}{\partial X^\mu\partial X^\nu}=0 \\
\end{array}
\right.\ \ ({\rm in\ the\ Onsager\ region})
\end{eqnarray}
\noindent where $\phi$ is a function, invariant under TCT. The solution of Eq.~(\ref{ts23}) is $\phi\equiv 0$ everywhere. 
The final expression of the affine connection for symmetric processes reads then
\begin{equation}\label{ts24}
\Gamma^\mu_{\alpha\beta}\!=\!
\begin{Bmatrix} 
\mu \\ \alpha\beta
\end{Bmatrix}
\!+\!\frac{1}{2\sigma}X^\mu\mathcal{O}(g_{\alpha\beta})\!-\!
\frac{1}{2(n+1)\sigma}\Bigl[
\delta^\mu_\alpha X^\nu\mathcal{O}(g_{\beta\nu})\!+\!\delta^\mu_\beta X^\nu\mathcal{O}(g_{\alpha\nu})\Bigr]
\end{equation}
\noindent The general case is given by
\begin{eqnarray}\label{ts25}
\ \!\Gamma^\mu_{\alpha\beta}=\!\!\!\!\!\!\!\!\!&&{\bar N}^{\mu\kappa}g_{\kappa\lambda}\begin{Bmatrix} 
\lambda \\ \alpha\beta
\end{Bmatrix}+\frac{{\bar N}^{\mu\kappa}}{2\sigma}X_\kappa\mathcal{O}(g_{\alpha\beta})
+\frac{{\bar N}^{\mu\kappa} }{2\sigma}X_\kappa X^\lambda\Bigl(\frac{\partial f_{\alpha\lambda}}{\partial X^\beta}+\frac{\partial f_{\beta\lambda}}{\partial X^\alpha}\Bigr)\nonumber\\
&&+\frac{{\bar N}^{\mu\kappa}}{2\sigma}f_{\kappa\varsigma}X^\varsigma X^\lambda\Bigl(\frac{\partial g_{\alpha\lambda}}{\partial X^\beta}+\frac{\partial g_{\beta\lambda}}{\partial X^\alpha}\Bigr)+\psi_\alpha\delta^\mu_\beta+\psi_\beta\delta^\mu_\alpha
\end{eqnarray}
\noindent 
where
\begin{eqnarray}\label{ts25a}
\psi_\nu=&&\!\!\!\!\!\!\!\!\!\!\!
-\frac{{\bar N}^{\eta\kappa}g_{\kappa\lambda}}{n+1}\begin{Bmatrix} 
\lambda \\ \eta\nu
\end{Bmatrix}
\!-\frac{{\bar N}^{\eta\kappa}X_\kappa}{2(n+1)\sigma}\mathcal{O}(g_{\nu\eta})\!-\!
\frac{{\bar N}^{\eta\kappa} }{2(n+1)\sigma}X_\kappa X^\lambda\Bigl(\frac{\partial f_{\eta\lambda}}{\partial X^\nu}+\frac{\partial f_{\nu\lambda}}{\partial X^\eta}\Bigr)
\nonumber\\
&&\!\!\!\!\!\!\!\!\!\!\!
-\frac{{\bar N}^{\eta\kappa}}{2(n+1)\sigma}f_{\kappa\varsigma}X^\varsigma X^\lambda\Bigl(\frac{\partial g_{\eta\lambda}}{\partial X^\nu}+\frac{\partial g_{\nu\lambda}}{\partial X^\eta}\Bigr)+\frac{1}{n+1}\frac{\partial\log\sqrt g}{\partial X^\nu}
\end{eqnarray}
Note that the thermodynamic space tends to reduce to a (thermodynamic) Riemannian space when ${\sigma}^{-1}\ll 1$. The following definitions are adopted:
\begin{itemize}
\item{ The space, covered by $n$ independent thermodynamic forces $X^\mu$, with metric tensor $g_{\mu\nu}$ and a linear single connection given by Eq.~(\ref{ts25}), may be referred to as {\it thermodynamic space} $T\!s$ (or, space of the thermodynamic forces).} 
\end{itemize}
In $T\!s$, the length of an arc is defined by the formula 
\begin{equation}\label{ts26}
L=\int_{\varsigma_1}^{\varsigma_2}\Bigl(g_{\mu\nu}\frac{ dX^\mu}{d\varsigma} \frac{ dX^\nu}{d\varsigma}\Bigr)^{1/2}d\varsigma
\end{equation}
\noindent The positive definiteness of matrix $g_{\mu\nu}$ ensures that $L\ge0$. Consider a coordinate system $X^\mu$, defining the thermodynamic space $T\!s$. 
\begin{itemize}
\item{All thermodynamic spaces obtained from $T\!s$ by a TCT transformation, may be called {\it iso-entropic spaces}.} 
\end{itemize}
In the TFT description, a thermodynamic configuration corresponds to a point in the thermodynamic space $T\!s$. The equilibrium state is the origin of the axes. Consider a thermodynamic system out of equilibrium, represented by a certain point, say $a$, in the thermodynamic space
\begin{itemize}
\item{A thermodynamic system is said {\it to relax} towards another point of the thermodynamic space, say $b$, if it moves from point $a$ to point $b$ following the shortest path~(\ref{ts3}), with the affine connection given in Eq.~(\ref{ts25}).}
\end{itemize}
\begin{itemize}
\item{ With Eq.~(\ref{ts25}), Eqs~(\ref{ts6}) may be called the {\it thermodynamic covariant differentiation} of a thermodynamic vector while Eqs.~(\ref{ts2}) and (\ref{ts5}) the {\it thermodynamic covariant differentiation along a curve} of a thermodynamic vector.}
\end{itemize}
\begin{itemize}
\item{With affine connection Eq.~(\ref{ts25}), $R^{\mu}_{\nu\lambda\kappa}$ may be called the {\it thermodynamic curvature tensor}.} 
\end{itemize}
\begin{itemize}
\item{The scalar $R$ obtained by contracting the thermodynamic tensor $R_{\nu\lambda}$ with $g^{\nu\lambda}$ (i.e. $R=R_{\nu\lambda}g^{\nu\lambda}$) may be called the {\it thermodynamic curvature scalar}.} 
\end{itemize}
\vskip 0.2truecm
\noindent{\bf The Principle of Least Action}
\vskip 0.2truecm
From expression (\ref{ts25}), the following mixed thermodynamic tensor of third order can be constructed
\begin{eqnarray}\label{pa1}
\Psi^\mu_{\alpha\beta}\equiv\!\!\!\!\!\!\!\!\!&&{\bar N}^{\mu\kappa}g_{\kappa\lambda}\begin{Bmatrix} 
\lambda \\ \alpha\beta
\end{Bmatrix}+\frac{{\bar N}^{\mu\kappa}}{2\sigma}X_\kappa\mathcal{O}(g_{\alpha\beta})
+\frac{{\bar N}^{\mu\kappa} }{2\sigma}X_\kappa X^\lambda\Bigl(\frac{\partial f_{\alpha\lambda}}{\partial X^\beta}+\frac{\partial f_{\beta\lambda}}{\partial X^\alpha}\Bigr)\nonumber\\
&&+\frac{{\bar N}^{\mu\kappa}}{2\sigma}f_{\kappa\varsigma}X^\varsigma X^\lambda\Bigl(\frac{\partial g_{\alpha\lambda}}{\partial X^\beta}+\frac{\partial g_{\beta\lambda}}{\partial X^\alpha}\Bigr)+\psi_\alpha\delta^\mu_\beta+\psi_\beta\delta^\mu_\alpha-\begin{Bmatrix} 
\mu \\ \alpha\beta
\end{Bmatrix}
\end{eqnarray}
\noindent This thermodynamic tensor satisfies the important identities
\begin{equation}\label{pa2}
\Psi^\alpha_{\alpha\beta}=\Psi^\beta_{\alpha\beta}=0
\end{equation}
\noindent Again, from $\Psi^\mu_{\alpha\beta}$ the mixed thermodynamic tensor of fifth order can be constructed
\begin{equation}\label{pa3}
S^{\mu\nu}_{\lambda\alpha\beta}\equiv 
\frac{1}{2}\Bigl(\Psi^\mu_{\beta\lambda}\delta^\nu_{\alpha}+
\Psi^\mu_{\alpha\lambda}\delta^\nu_{\beta}+
\Psi^\nu_{\beta\lambda}\delta^\mu_{\alpha}+
\Psi^\nu_{\alpha\lambda}\delta^\mu_{\beta}
-\Psi^\mu_{\alpha\beta}\delta^\nu_\lambda
-\Psi^\nu_{\alpha\beta}\delta^\mu_\lambda\Bigr)
\end{equation}
\noindent By contraction, a thermodynamic tensor of third order, a thermodynamic vector and a thermodynamic scalar can be formed as follows
\begin{eqnarray}\label{pa4}
&&\!\!\!\!\!\!\!\!\!\!\!\!\!\!\!\!\!\!\!\!
S^{\mu\nu}_{\lambda}\equiv S^{\mu\nu}_{\lambda\alpha\beta}g^{\alpha\beta}=\Psi^\mu_{\lambda\alpha}g^{\nu\alpha}+
\Psi^\nu_{\lambda\alpha}g^{\mu\alpha}-\frac{1}{2}\Psi^\mu_{\alpha\beta}g^{\alpha\beta}\delta^\nu_\lambda-\frac{1}{2}\Psi^\nu_{\alpha\beta}g^{\alpha\beta}\delta^\mu_\lambda\nonumber\\
&&\!\!\!\!\!\!\!\!\!\!\!\!\!\!\!\!\!\!\!\!
S^{\mu}\equiv S^{\mu\lambda}_{\lambda}=\frac{1-n}{2}\Psi^\mu_{\alpha\beta}g^{\alpha\beta}\\
&&\!\!\!\!\!\!\!\!\!\!\!\!\!\!\!\!\!\!\!\!
S\equiv S^{\mu\nu}_{\lambda}\Psi^{\lambda}_{\mu\nu}=2\Psi^\kappa_{\lambda\mu}\Psi^\lambda_{\kappa\nu}g^{\mu\nu}\nonumber
\end{eqnarray}
\noindent The following postulate is now introduced:

\noindent {\it There exists a thermodynamic action $I$, scalar under $TCT$, which is stationary with respect to arbitrary variations in the transport coefficients and the affine connection}. 

\noindent This action, scalar under $TCT$, constructed from the transport coefficients and their first and second derivatives, should have linear second derivatives. In addition, it should be stationary when the affine connection takes the expression given in Eq.~(\ref{ts25}). The only action satisfying these requirements is 
\begin{equation}\label{pa5}
I=\int\Bigl[ R_{\mu\nu}-(\Gamma^\lambda_{\alpha\beta}-
{\tilde\Gamma}^\lambda_{\alpha\beta})S^{\alpha\beta}_{\lambda\mu\nu}
\Bigr]g^{\mu\nu}\sqrt{g}\ \! {d^{}}^n\!X
\end{equation}
\noindent where ${d^{}}^n\!X$ denotes an infinitesimal volume element in $T\!s$ and ${\tilde\Gamma}^\kappa_{\mu\nu}$ is the expression given in Eq.~(\ref{ts25}) i.e., ${\tilde\Gamma}^\kappa_{\mu\nu}=\Psi^\kappa_{\mu\nu}+\Bigl\{{}^{\ \!\kappa}_{\mu\nu}\Bigr\}$. To avoid misunderstanding, while it is correct to mention that this postulate affirms the possibility of deriving the thermodynamic field equations by a variational principle it does not state that the expressions and theorems obtained from the solutions of the thermodynamic field equations can also be derived by a variational principle. In particular the well-known {\it Universal Criterion of Evolution} established by Glansdorff-Prigogine {\it can not} be derived by a variational principle (see also section \ref{ttse}).
\vskip 0.2truecm
\noindent{\bf The Thermodynamic Field Equations}
\vskip 0.2truecm
As a first step, the transport coefficients and the affine connection are subjected to infinitesimal variations i.e., $g_{\mu\nu}\rightarrow g_{\mu\nu}+\delta g_{\mu\nu}$, $f_{\mu\nu}\rightarrow f_{\mu\nu}+\delta f_{\mu\nu}$ and $\Gamma^\kappa_{\mu\nu}\rightarrow \Gamma^\kappa_{\mu\nu}+\delta \Gamma^\kappa_{\mu\nu}$, where $\delta g_{\mu\nu}$, $\delta f_{\mu\nu}$ and $\delta \Gamma^\kappa_{\mu\nu}$ are arbitrary, except that they are required to vanish as $\mid X^\mu\mid\rightarrow\infty$. By imposing that the action (\ref{pa5}) is stationary with respect to arbitrary variations in $g_{\mu\nu}$, $f_{\mu\nu}$ and $\Gamma^\kappa_{\mu\nu}$, the thermodynamic field equations (i.e., the equations for the transport coefficients and the affine connection) are derived. The results are, respectively (see appendix \ref{appeq})
\begin{eqnarray}\label{tfe1}
&& R_{\mu\nu}-\frac{1}{2}g_{\mu\nu}R=-S^{\alpha\beta}_\lambda\frac{\delta{\tilde\Gamma}^\lambda_{\alpha\beta}}{\delta g^{\mu\nu}}\nonumber\\
&&S^{\alpha\beta}_\lambda\frac{\delta{\tilde\Gamma}^\lambda_{\alpha\beta}}{\delta f^{\mu\nu}}=0\\
&&g_{\mu\nu\mid\lambda}=-\Psi^\alpha_{\mu\lambda}g_{\alpha\nu}-\Psi^\alpha_{\nu\lambda}g_{\alpha\mu}
\nonumber
\end{eqnarray}
\noindent where the variations of the affine connection (\ref{ts25}) with respect to the transport coefficients appear in the first two equations. From the first equation of Eqs~(\ref{tfe1}), the expression for the thermodynamic curvature scalar is given as
\begin{equation}\label{tfe2}
R=\frac{2}{n-2}g^{\mu\nu}S^{\alpha\beta}_\lambda \frac{\delta{\tilde\Gamma}^\lambda_{\alpha\beta}}{\delta g^{\mu\nu}}
\end{equation}
\noindent The third equation of Eqs~(\ref{tfe1}) can be re-written as
\begin{equation}\label{tfe2}
g_{\mu\nu,\lambda}
-\Gamma^\alpha_{\mu\lambda}g_{\alpha\nu}-\Gamma^\alpha_{\nu\lambda}g_{\alpha\mu}
=-\Psi^\alpha_{\mu\lambda}g_{\alpha\nu}-\Psi^\alpha_{\nu\lambda}g_{\alpha\mu}
\end{equation}
\noindent where the comma $(,)$ denotes partial differentiation. Adding to this equation the same equation with $\mu$ and $\lambda$ interchanged and subtracting the same equation with $\nu$ and $\lambda$ interchanged gives
\begin{equation}\label{tfe2}
g_{\mu\nu,\lambda}+g_{\lambda\nu,\mu}-g_{\mu\lambda,\nu}=2g_{\alpha\nu}\Gamma^\alpha_{\lambda\mu}-2g_{\alpha\nu}\Psi^\alpha_{\lambda\mu}
\end{equation}
\noindent or 
\begin{equation}\label{tfe3}
\Gamma^\kappa_{\lambda\mu}=\Bigl\{{}^{\ \!\kappa}_{\lambda\mu}\Bigr\}+\Psi^\kappa_{\lambda\mu}={\tilde\Gamma}^\kappa_{\lambda\mu}
\end{equation}
\noindent For $a-a$ processes, close to the Onsager region, it holds that
\begin{eqnarray}\label{tfe5}
\!\!\!\!\!\!\!\!\!\!\!\!\!\!\!\!&&g_{\mu\nu}=L_{\mu\nu}+h_{\mu\nu}+O(\epsilon^2)\nonumber\\
\!\!\!\!\!\!\!\!\!\!\!\!\!\!\!\!&&\lambda_\sigma=O(\epsilon)\qquad {\rm with}\quad \epsilon=Max\Big\{\frac{\mid Eigenvalues[g_{\mu\nu}-L_{\mu\nu}]\mid}{Eigenvalues [L_{\mu\nu}]}\Big\}\ll 1
\end{eqnarray}
\noindent where $\lambda_\sigma\equiv 1/\sigma$ and $h_{\mu\nu}$ are small variations with respect to Onsager's coefficients. In this region, Eq.~(\ref{pa5}) is stationary for arbitrary variations of $h_{\mu\nu}$ and $\Gamma^\kappa_{\mu\nu}$. It can be shown that \cite{sonnino}
\begin{eqnarray}\label{tfe6}
\!\!\!\!\!\!\!\!\!\!\!\!&&L^{\lambda\kappa}\frac{\partial^2 h_{\mu\nu}}{\partial X^{\lambda}\partial X^{\kappa}}+
L^{\lambda\kappa}\frac{\partial^2 h_{\lambda\kappa}}{\partial X^{\mu}\partial X^{\nu}}-
L^{\lambda\kappa}\frac{\partial^2 h_{\lambda\nu}}{\partial X^{\kappa}\partial X^{\mu}}-
L^{\lambda\kappa}\frac{\partial^2 h_{\lambda\mu}}{\partial X^{\kappa}\partial X^{\nu}}=0+O(\epsilon^2)\nonumber\\
\!\!\!\!\!\!\!\!\!\!\!\!&&g_{\mu\nu;\lambda}=0+O(\epsilon^2)\qquad {\rm or} \quad\Gamma_{\mu\nu}^\kappa=\frac{1}{2}L^{\kappa\eta}(h_{\mu\eta,\nu}+h_{\nu\eta,\mu}-h_{\mu\nu,\eta})+O(\epsilon^2)
\end{eqnarray}
\noindent where the semicolon indicates the covariant derivative performed with the Christoffel symbols. Eqs~(\ref{tfe5}) should be solved with the appropriate gauge-choice and boundary conditions.

\noindent The validity of Eqs~(\ref{tfe6}) has been largely tested by analyzing several symmetric processes, such as the thermoelectric effect and the unimolecular triangular chemical reactions \cite{sonnino}. More recently, these equations have been also used to study transport processes in magnetically confined plasmas. In all examined examples, the theoretical results of the TFT are in line with experiments. It is worthwhile mentioning that, for transport processes in tokamak plasmas, the predictions of the TFT for radial energy and matter fluxes are much closer to the experimental data than the neoclassical theory (based on linearized Boltzmann's equation), which fails with a factor $10^3\div10^4$ \cite{balescu2} and \cite{sonnino3}. 
\vskip 0.2truecm
\noindent{\bf Some Remarks on Spatially Extended Thermodynamic Systems}
\vskip 0.2truecm
The macroscopic description of thermodynamic systems gives rise to state variables that depend continuously on space coordinates. In this case, the thermodynamic forces possess an infinity associated to each point of the space coordinates. The system may be subdivided into $N$ cells ($N$x$N$x$N$ in three dimensions), each of which labeled by a wave-number ${\bf k}$, and we follow their relaxation. Without loss of generality, we consider a thermodynamic system confined in a rectangular box with sizes $l_x$, $l_y$ and $l_z$. Denoting with ${\bf k}$ the wave-number 
\begin{equation}\label{sets1}
{\bf k}=2\pi\bigl(\frac{n_x}{l_x}, \frac{n_y}{l_y}, \frac{n_z}{l_z}\bigr)\quad{\rm with}\quad
\left\{ \begin{array}{ll}
n_x= 0,\pm1,\cdots\pm N_x\\
n_y= 0,\pm1,\cdots\pm N_y\\
n_z= 0,\pm1,\cdots\pm N_z
\end{array}
\right.
\end{equation}
\noindent the fluxes and forces, developed in (spatial) Fourier's series, read
\begin{eqnarray}\label{sets2}
&&\!\!\!\!\!\!\!\!\!\!\!\!
{\mathcal J}_\mu({\bf r},t)=\!\!\!\!\sum_{{\bf n}=-{\bf N}}^{\bf N}{\hat J}_{\mu ({\bf k})}(t)\exp(i{\bf k}\cdot{\bf r})\nonumber\\
&&\!\!\!\!\!\!\!\!\!\!\!\!
{\mathcal X}^\mu({\bf r},t)=\!\!\!\!\sum_{{\bf n}'=-{\bf N}}^{\bf N}{\hat X}^\mu_{({\bf k}')}(t)\exp(i{\bf k}'\cdot{\bf r})
\end{eqnarray}
\noindent where, for brevity, ${\bf n}$ and ${\bf N}$ stand for ${\bf n}=(n_x,n_y,n_z)$ and ${\bf N}=(N_x,N_y,N_z)$, respectively. The Fourier coefficients are given by 
\begin{eqnarray}\label{sets3}
&&\!\!\!\!\!\!\!\!\!\!\!\!
{\hat J}_{\mu ({\bf k})}(t)=\frac{1}{\Omega}\int_\Omega {\mathcal J}_\mu({\bf r},t)
\exp(-i{\bf k}\cdot{\bf r})dv\nonumber\\
&&\!\!\!\!\!\!\!\!\!\!\!\!
{\hat X}^\mu_{({\bf k}')}(t)=\frac{1}{\Omega}\int_\Omega{\mathcal X}^\mu({\bf r},t)
\exp(-i{\bf k}'\cdot{\bf r})dv
\end{eqnarray}
\noindent In particular, the contributions at the thermodynamic limit (i.e., for ${\bf k}\rightarrow 0$) are expressed as 
\begin{eqnarray}\label{sets4}
&&\!\!\!\!\!\!\!\!\!\!\!\!
{\hat J}_{\mu (0)}(t)=\frac{1}{\Omega}\int_\Omega {\mathcal J}_\mu({\bf r},t)
dv=J_\mu(t)\nonumber\\
&&\!\!\!\!\!\!\!\!\!\!\!\!
{\hat X}^\mu_{(0)}(t)=\frac{1}{\Omega}\int_\Omega{\mathcal X}^\mu({\bf r},t)
dv=X^\mu(t)
\end{eqnarray}
\noindent The entropy production and the transport relations take, respectively, the form
\begin{eqnarray}\label{sets5}
&&\!\!\!\!\!\!\!\!\!\!\!\!
 {\mathcal\sigma}({\bf r},t)={\mathcal J}_\mu({\bf r},t){\mathcal X}^\mu({\bf r},t)\geq0
\nonumber\\
&&\!\!\!\!\!\!\!\!\!\!\!\!
{\mathcal J}_\mu({\bf r},t)={\mathcal\tau}_{\mu\nu}({\bf r},t){\mathcal X}^\nu({\bf r},t)
\end{eqnarray}
\noindent Considering that 
\begin{eqnarray}\label{sets6}
&&\!\!\!\!\!\!\!\!\!\!\!\!\!\!\!\!\!\!\!\!\!\!\!\!\!\!\!\!\!\!\!\!\!\!\!\!
\int_0^{l_x}\!\!\int_0^{l_y}\!\!\int_0^{l_z}\!\!\exp[i({\bf k}+{\bf k}')\cdot{\bf r}]dv =\Omega\ \delta_{{\bf k}+{\bf k}',0}\qquad {\rm with}\\
&&\!\!\!\!\!\!\!\!\!\!\!\!\!\!\!\!\!\!\!\!\!\!\!\!\!\!\!\!\!\!\!\!\!\!\!\!
\delta_{{\bf k}+{\bf k}',0}=
\left\{ \begin{array}{ll}
0& \mbox{if $ {\bf k}+{\bf k}'\neq 0$}\\
1& \mbox{if $ {\bf k}+{\bf k}'= 0$}
\end{array}\qquad\qquad {\rm and}\qquad\!\! \Omega=l_xl_yl_z
\right.\nonumber
\end{eqnarray}
\noindent from the first equation of Eq.~(\ref{sets5}) we also find
\begin{equation}\label{sets7}
\int_\Omega
{\mathcal J}_\mu({\bf r},t){\mathcal X}^\mu({\bf r},t)\ dv=\Omega\Bigl({\hat J}_{\mu (0)}(t){\hat X}^\mu_{(0)}(t)+
\sum_{{\bf k}\neq0}
{\hat J}_{\mu ({\bf k})}(t){\hat  X}^\mu_{({\bf -k})}(t)\Bigr)\geq0
\end{equation}
\noindent On the other hand, we have
\begin{equation}\label{sets8}
{\hat J}_{\mu (0)}(t)={\hat \tau}_{\mu\nu(0)}(t){\hat X}^\nu_{(0)}(t)+\sum_{{\bf k}\neq0}{\hat \tau}_{\mu\nu({\bf k})}(t){\hat X}^\nu_{({\bf -k})}(t)
\end{equation}
\noindent where
\begin{equation}\label{sets7a}
{\hat \tau}_{\mu\nu({\bf k})}(t)=\frac{1}{\Omega}\int_\Omega \tau_{\mu\nu}({\bf r},t)
\exp(-i{\bf k}\cdot{\bf r})dv
\end{equation}
\noindent Eq.~(\ref{sets7}) can then be brought into the form
\begin{eqnarray}\label{sets9}
&&\!\!\!\!\!\!\!\!\!\!\!\!\!\!\!\!\!\!\!
\int_\Omega
\sigma\ dv=\Omega{\hat g}_{\mu\nu(0)}(t){\hat X}^\mu_{(0)}(t){\hat X}^\nu_{(0)}(t)\nonumber\\
&&\quad\ 
+\Omega
\sum_{{\bf k}\neq0}\Bigl({\hat \tau}_{\mu\nu({\bf k})}(t){\hat X}^\nu_{({\bf -k})}(t){\hat X}^\nu_{(0)}(t)
+{\hat J}_{\mu ({\bf k})}(t){\hat  X}^\mu_{({\bf -k})}(t)\Bigr)\geq0
\end{eqnarray}
\noindent where
\begin{eqnarray}\label{sets9a}
&&{\hat g}_{\mu\nu({\bf k})}(t)=\frac{1}{\Omega}\int_\Omega {\mathcal G}_{\mu\nu}({\bf r},t)
\exp(-i{\bf k}\cdot{\bf r})dv\qquad{\rm with}\nonumber\\
&&{\mathcal G}_{\mu\nu}({\bf r},t)\equiv\frac{1}{2}[\tau_{\mu\nu}({\bf r},t)+\tau_{\nu\mu}({\bf r},t)]
\end{eqnarray}
\noindent In Eq.~(\ref{sets9}), the first term is the contribution at the thermodynamic limit whereas the second expression reflects the interactions between the ${\bf k}$-cell and the other cells. In a relaxation process, contributions from different wave-numbers are negligible with respect to those with same wave-numbers and, hence, we finally obtain 
\begin{equation}\label{sets10}
\int_\Omega
\sigma\ dv\simeq\Omega
{\hat g}_{\mu\nu(0)}(t){\hat X}^\mu_{(0)}(t){\hat X}^\nu_{(0)}(t)>0\qquad\forall\ {\hat X}^\mu_{(0)}(t)\ ({\rm and}\ \sigma\neq0)
\end{equation}
\noindent Last inequality is satisfied {\it for any} ${\hat X}^\mu_{(0)}(t)$ if, and only if
\begin{equation}\label{sets11}
{\hat g}_{\mu\nu(0)}(t)=\frac{1}{\Omega}\int_\Omega{\mathcal G}_{\mu\nu}({\bf r},t)\ dv=g_{\mu\nu}(t)
\end{equation}
\noindent is a {\it positive definite matrix}. Therefore, for spatially extended thermodynamic systems, we replace $X^\mu(t)\rightarrow X^\mu_{({\bf k})}(t)$ and $\tau_{\mu\nu} (t)\rightarrow \tau_{\mu\nu({\bf k})} (t)$. Under these conditions, Eqs~(\ref{tfe1}) determine the nonlinear corrections to the Onsager coefficients while Eqs~(\ref{ts2}) and Eqs~(\ref{ts6}), with affine connection Eq.~(\ref{ts25}), are the thermodynamic covariant differentiation along a curve and the thermodynamic covariant differentiation of a thermodynamic vector, respectively. 

\vskip 0.2truecm
\noindent{\bf The Privileged Thermodynamic Coordinate System}
\vskip 0.2truecm
By definition, a thermodynamic coordinate system is a set of coordinates defined so that the expression of the entropy production takes the form of Eq.~(\ref{tr1}). Once a particular set of thermodynamic coordinates is determined, the other sets of coordinates are linked to the first one through a TCT [see Eqs~(\ref{tr2})]. The simplest way to determine a particular set a coordinate transformation is to quote the entropy balance equation
\begin{equation}\label{ptcs1}
\frac{\partial \rho s}{\partial t}+{\bf\nabla}\cdot{\bf J}_s=\sigma
\end{equation}
 \noindent where $\rho s$ is the local total entropy per unit volume and ${\bf J}_s$ is the entropy flux. Let us consider, as an example, a thermodynamic system confined in a rectangular box where chemical reactions, diffusion of matter, macroscopic motion of the volume element (convection) and heat current take place simultaneously. The entropy flux and the entropy production read \cite{prigogine2}, \cite{fitts}
 \begin{eqnarray}\label{pct2}
&&\!\!\!\!\!\!\!\!\!\!\!\!\!\!\!\!\!\!\!\!
 {\bf J}_s=\frac{1}{T}({\bf J}_q-\sum_i {\bf J}_i\mu_i)+\sum_i\rho_i v_is_i\nonumber\\
&&\!\!\!\!\!\!\!\!\!\!\!\!\!\!\!\!\!\!\!\!
\sigma={\bf J}_q\!\cdot\!{\bf\nabla}\frac{1}{T}\!-\!\frac{1}{T}\!\sum_i{\bf J}_i\!\cdot\!\Bigl[{\bf\nabla}\Bigl(\frac{\mu_i}{T}\Bigl)\!-\!{\bf F}_i\Bigr]\!+\!\sum_i\frac{w_iA_i}{T}\!-\!\frac{1}{T}\!\sum_{ij}\Pi_{ij}\partial_{{\bf r}_i}v_j\geq0\\
\end{eqnarray}
 \noindent where $\mu_i$, $\rho_i s_i$ and $A_i$ are the chemical potential, the local entropy and the affinity of species "$i$", respectively. Moreover, ${\bf F}_i$ indicates the external force per unit mass acting on $"i"$, $\Pi_{ij}$ the components of the dissipative part of the pressure tensor ${\mathcal M}_{ij}$ (${\mathcal M}_{ij}=p\delta_{ij}+\Pi_{ij}$; $p$ is the hydrostatic pressure) and $v_j$ is the component of the hydrodynamic velocity (see, for example, ref. \cite{vidal}). The set of thermodynamic coordinates is given as
 \begin{equation}\label{ptcs3}
\Bigl\{{\bf\nabla}\frac{1}{T};\ -\frac{1}{T}\Bigl[{\bf\nabla}\Bigl(\frac{\mu_i}{T}\Bigl)-{\bf F}_i\Bigr];\ \frac{A_i}{T};\ -\frac{1}{T}\partial_{{\bf r}_i}v_j\Bigr\}
\end{equation}
 \noindent For this particular example, this set may be referred to as the {\it privileged thermodynamic coordinate system}. Other examples of privileged thermodynamic coordinate system, concerning magnetically confined plasmas, can be found in refs \cite{balescu2}, \cite{sonnino3} and \cite{balescu1}.
\vskip 0.2truecm
\section{Thermodynamic Theorems for Systems out of Equilibrium}\label{ttse}
\vskip 0.2truecm
In 1945-1947, Prigogine proved the minimum entropy production theorem \cite{prigogine}, which concerns the relaxation of thermodynamic systems near equilibrium. This theorem states that:
\vskip 0.2truecm
\noindent {\bf Minimum Entropy Production Theorem} (MEPT)

{\it For $a-a$ processes, a thermodynamic system, near equilibrium, relaxes to a steady-state $X_s$ in such a way that the inequality}
\begin{equation}\label{tt1}
\frac{d\sigma}{dt}\le0
\end{equation}
\noindent {\it is satisfied throughout the evolution and is only saturated at $X_s$}.

\noindent The minimum entropy production theorem is generally not satisfied far from equilibrium. However, P. Glansdorff and I. Prigogine demonstrated in 1964 that a similar theorem continues to hold for any relaxation to a steady-state, which reads \cite{prigogine1}
\vskip 0.2truecm
\noindent {\bf Universal Criterion of Evolution} (UCE)

{\it When the thermodynamic forces and conjugate flows are related by a generic asymmetric tensor, regardless of the type of processes, for time-independent boundary conditions a thermodynamic system, even in strong non-equilibrium conditions, relaxes towards a steady-state in such a way that the following universal criterion of evolution is satisfied}:
\begin{equation}\label{tt2}
{\mathcal P}\equiv J_\mu\frac{dX^\mu}{dt}\le0
\end{equation}
\noindent {\it This inequality is only saturated at $X_s$}. 

\noindent For $a-a$ processes, the UCE reduces to the MEPT in the Onsager region. Again, Glansdorff and Prigogine demonstrated this theorem using a purely thermodynamical approach. In this section we shall see that if the system relaxes towards a steady-state along the shortest path then the Universal Criterion of Evolution is automatically satisfied.

\noindent By definition, a necessary and sufficient condition for a curve to be the shortest path is that it satisfies the differential equation 
\begin{equation}\label{tt2a}
\frac{d^2X^\mu}{dt^2}+\Gamma^\mu_{\alpha\beta}\frac{dX^\alpha}{dt} \frac{dX^\beta}{dt}=\varphi(t)\frac{dX^\mu}{dt}
\end{equation}
\noindent where $\varphi(t)$ is a determined function of time. If we define a parameter $\varrho$ by
\begin{equation}\label{tt2b}
\frac{d\varrho}{dt}=c\exp\int\varphi^{*} dt\qquad{\rm with}\quad \varphi^{*}=\varphi-2\psi_\nu\frac{dX^\nu}{dt}
\end{equation}
\noindent where $c$ is an arbitrary constant and $\psi_\nu$ the projective covariant vector, Eq.~(\ref{tt2a}) reduces to Eq.~(\ref{ts3}) with $\Gamma^\mu_{\alpha\beta}$ given by Eq.~(\ref{ts9}). Parameter $\varrho$ is not the affine parameter $s$ of the shortest path. The relation between these two parameters is 
\begin{equation}\label{tt2c}
\varrho=b\int\exp\Bigl(-2\int\psi_\nu dX^\nu\Bigr)ds
\end{equation}
\noindent where $b$ is an arbitrary constant. Eq.~(\ref{tt2b}) allows us to choose the parameter $\varrho$ in such a way that it increases monotonically as the thermodynamic system evolves in time. In this case, $c$ is a positive constant and, without loss of generality, we can set $c=1$. Parameter $\varrho$ can also be chosen so that it vanishes when the thermodynamic system begins to evolve and it takes the (positive) value, say ${\bar l}$, when the system reaches the steady-state. Multiplying Eq.~(\ref{ts3}) with the flows $J_\mu$ and contracting, we obtain
\begin{equation}\label{tt3}
J_\mu\frac{d^2X^\mu}{d\varrho^2}+J_{\mu}\Gamma^\mu_{\alpha\beta}\frac{dX^\alpha}{d\varrho} \frac{dX^\beta}{d\varrho} =0
\end{equation}
\noindent However 
\begin{equation}\label{tt4}
 J_\mu\frac{d^2X^\mu}{d\varrho^2}=\frac{d{\tilde P}}{d\varrho}-\Bigl(\frac{d\varsigma}{d\varrho}\Bigr)^2-
 \frac{dX^\alpha}{d\varrho} \frac{dX^\beta}{d\varrho} 
  X^\lambda\frac{\partial g_{\alpha\lambda}}{\partial X^\beta}- \frac{dX^\alpha}{d\varrho} \frac{dX^\beta}{d\varrho}X^\lambda\frac{\partial f_{\alpha\lambda}}{\partial X^\beta}
\end{equation}
\noindent where ${\tilde P}=J_\mu \frac{dX^\mu}{d\varrho} $ and after taking into account the identities $f_{\mu\nu}\frac{dX^\mu}{d\varrho} \frac{dX^\nu}{d\varrho} =0$ and $g_{\mu\nu}\frac{dX^\mu}{d\varsigma} \frac{dX^\nu}{d\varsigma} =1$. In addition, recalling Eq.~(\ref{ts13}) and the relations $X_\mu X^\mu=\sigma$ and $f_{\mu\nu}X^\mu X^\nu=0$, it can be shown that
\begin{equation}\label{tt5}
J_{\mu}\Gamma^\mu_{\alpha\beta} \frac{dX^\alpha}{d\varrho} \frac{dX^\beta}{d\varrho}= \frac{dX^\alpha}{d\varrho} \frac{dX^\beta}{d\varrho} X^\lambda\frac{\partial g_{\alpha\lambda}}{\partial X^\beta}+ \frac{dX^\alpha}{d\varrho} \frac{dX^\beta}{d\varrho} X^\lambda\frac{\partial f_{\alpha\lambda}}{\partial X^\beta}
\end{equation}
\noindent Summing Eq.~(\ref{tt4}) with Eq.~(\ref{tt5}) and considering Eq.~(\ref{tt3}), gives
\begin{equation}\label{tt6}
\frac{d {\tilde P}}{d\varrho}=\Bigl(\frac{d\varsigma}{d\varrho}\Bigr)^2
\end{equation}
\noindent Integrating Eq.~(\ref{tt6})] from the initial condition to the steady-state, we find
\begin{equation}\label{tt6a}
{\tilde P}(X_s)-{\tilde P}=\int\Bigl(\frac{d\varsigma}{d\varrho}\Bigr)^2d\varrho\ge0
\end{equation}
\noindent From Eq.~(\ref{o2}) we have ${\tilde P}(X_s)=P(X_s)d\varsigma/d\varrho=0$, so we finally obtain
\begin{equation}\label{tt7}
{\tilde P}=J_\mu\frac{dX^\mu}{d\varrho}=-\int\Bigl(\frac{d\varsigma}{d\varrho}\Bigr)^2d\varrho\le0
\end{equation}
\noindent where the inequality is only saturated at the steady-state. Recalling Eq.~(\ref{tt2b}), the inequality established by the UCE can be derived
\begin{equation}\label{tt8}
{\mathcal P}={\tilde P}\frac{d\varrho}{dt}=J_\mu\frac{dX^\mu}{d\varrho}\Bigl(\exp\int\varphi^{*} dt\Bigr)\le0
\end{equation}
\noindent Eq.~(\ref{tt6}) can be re-written as 
\begin{equation}\label{tt8a}
\frac{d}{d\varsigma}\Bigl[\Bigl(\frac{d\varsigma}{d\varrho}\Bigr)P\Bigr]=\Bigl(\frac{d\varsigma}{d\varrho}\Bigr)
\end{equation}
\noindent This equation generalizes Eq.~(\ref{i12}), which was valid only in the near equilibrium region (notice that, in the linear region, $d\varsigma/d\varrho=1/b$). Integrating Eq.~(\ref{tt8a}), the expression of the dissipative quantity $P$ is derived
\begin{equation}\label{tt8b}
P=-\Bigl(\frac{d\varrho}{d\varsigma}\Bigr)\int_\varsigma^l\Bigl(\frac{d\varsigma'}{d\varrho}\Bigr)d\varsigma'=-{\Bigl(g_{\mu\nu}\frac{dX^\mu}{d\varrho}\frac{dX^\nu}{d\varrho}\Bigr)}^{\!\!-1/2}\!\!\!\!\int_\varsigma^l\!\!{\Bigl(g_{\mu\nu}\frac{dX^\mu}{d\varrho}\frac{dX^\nu}{d\varrho}\Bigr)}^{\!\!1/2}\!\!\!d\varsigma'
\end{equation}
\noindent On the right, it is understood that the $X$'s are expressed in terms of $\varrho(\varsigma)$. Eq.~(\ref{tt8b}) generalizes Eq.~(\ref{i10}), which was valid only in the linear region. For $a-a$ processes in the Onsager region, Eq.~(\ref{tt8}) implies the validity of the inequality (\ref{tt1}). Indeed, Eq.~(\ref{ts4}) gives
\begin{equation}\label{tt9}
\frac{\delta\sigma}{\delta\varrho}=\frac{d\sigma}{d\varrho}=J_\mu\frac{\delta X^\mu}{\delta\varrho}+X^\mu\frac{\delta  J_\mu}{d\varrho}=2J_\mu\frac{\delta X^\mu}{\delta\varrho}+X^\mu X^\nu\frac{\delta L_{\mu\nu}}{\delta\varrho}
\end{equation}
\noindent In the linear region, the coefficients of the affine connection vanish. Eq.~(\ref{tt9}) is simplified reducing to
\begin{equation}\label{tt10}
\frac{d\sigma}{dt}=\frac{d\sigma}{d\varrho}\frac{d\varrho}{dt}=2\Bigl(J_\mu\frac{d X^\mu}{d\varrho}\frac{d\varrho}{dt}\Bigr)=2{\mathcal P}\le0
\end{equation}
\noindent where the inequality is saturated only at the steady state. Let us now consider spatially extended thermodynamic systems. In space-dependent systems, the dissipative quantity should be expressed in the integral form
\begin{equation}\label{tt11}
\mathcal{P}= \int_\Omega {\mathcal J}_\mu({\bf r},t) d_t{\mathcal X}^\mu({\bf r},t)\ dv 
\end{equation}
\noindent where $d_t{\mathcal X}^\mu\equiv d{\mathcal X}^\mu/dt$. In terms of wave-vectors ${\bf k}$, Eq.~(\ref{tt11}) can be brought into the form
\begin{equation}\label{tt15}
\mathcal{P}= \Omega \Bigl({\hat J}_{\mu (0)}(t){d_t{\hat X}}^\mu_{(0)}(t)+\sum_{{\bf k}\neq0}
{\hat J}_{\mu ({\bf k})}(t)d_t{\hat  X}^\mu_{({\bf -k})}(t)\Bigr)
\end{equation}
\noindent where Eq.~(\ref{sets6}) has been taken into account. As already mentioned in section \ref{isoef}, in a relaxation process, contributions from different wave-numbers are negligible with respect to those with same wave-numbers and, hence, we finally obtain 
\begin{equation}\label{tt16}
\mathcal{P}\simeq \Omega J_{\mu}(t)d_t X^\mu(t)\leq 0
\end{equation}
\noindent
where Eq.~(\ref{tt8}) has been taken into account. It is therefore proven that the Universal Criterion of Evolution is automatically satisfied if the system relaxes along the shortest path. Indeed it would be more exact to say: the affine connection, given in Eq.~(\ref{ts9}), has been constructed in such a way that the UCE is satisfied without imposing {\it any} restrictions on the transport coefficients (i.e., on matrices $g_{\mu\nu}$ and $f_{\mu\nu}$). In addition, analogously to Christoffel's symbols, the elements of the affine connection are constructed from matrices $g_{\mu\nu}$ and $f_{\mu\nu}$ and their first derivatives in such a way that all coefficients vanish in the Onsager region. Eq.~(\ref{ts9}) provides the simplest expression satisfying these requirements.
\vskip 0.2truecm
\noindent {\bf The Minimum Rate of Dissipation Principle} (MRDP)

In ref.\cite{sonnino1} the validity of the following theorem is shown: 

\noindent {\it The generally covariant part of the Glansdorff-Prigogine quantity is always negative and is locally minimized when the evolution of a system traces out a geodesic in the space of thermodynamic configurations}.

\noindent It is important to stress that this theorem does not refer to the Glansdorff-Prigogine expression reported in Eq.~(\ref{tt2}) but only to its {\it generally covariant part}. Moreover, it concerns the evolution of a system in the space of thermodynamic configurations and {\it not} in the thermodynamic space. One could consider the possibility that the shortest path in the thermodynamic space is an extremal for the functional
\begin{equation}\label{tt17}
\int_{\varsigma_1}^{\varsigma_2}J_\mu{\dot X}^\mu d\varsigma
\end{equation}
\noindent The answer is negative. Indeed, a curve is an extremal for functional Eq.~(\ref{tt17}) if, and only if, it satisfies Euler's equations\footnote{Notice that $J_{\nu,\mu}-J_{\mu,\nu}$ is a thermodynamic tensor of second order. \label{Euler}}
\begin{equation}\label{tt18}
{\dot X}^\nu\Bigl(\frac{\partial J_\nu}{\partial X^\mu}-\frac{\partial J_\mu}{\partial X^\nu}\Bigr)=0
\end{equation}
\noindent 
As it can be easily checked, this extremal coincides with the shortest path if
\begin{eqnarray}\label{tt19}
&&\!\!\!\!\!\!\!\!\!\!\!\!\!\!\!\!  
\frac{1}{2}\Bigl(\frac{M_{\mu\alpha}}{\partial X^\beta}+\frac{M_{\mu\beta}}{\partial X^\alpha}\Bigr)-\Gamma^{\kappa}_{\alpha\beta}M_{\mu\kappa}=0\qquad{\rm where}\\
&&\!\!\!\!\!\!\!\!\!\!\!\!\!\!\!\!
M_{\mu\nu}\equiv J_{\nu,\mu}-J_{\mu,\nu}=2f_{\nu\mu}+X^\kappa(g_{\nu\kappa,\mu}-g_{\mu\kappa,\nu})+X^\kappa(f_{\nu\kappa,\mu}-f_{\mu\kappa,\nu})\nonumber
\end{eqnarray}
\noindent and $\Gamma^{\kappa}_{\alpha\beta}$ given in Eq.~(\ref{ts25}). However, Eqs.~(\ref{tt19}) are $n^2(n+1)/2$ equations for $n^2$ variables (the transport coefficients) and, in general, for $n\neq 1$, they do not admit solutions. We have thus another proof that the Universal Criterion of Evolution can not be derived from a variational principle.

\vskip 0.2truecm
\section{Conclusions and Limit of Validity of the Approach}\label{cs}
\vskip 0.2truecm
\noindent The main purpose of this paper is to present a new formulation of the thermodynamic field theory (TFT) where one of the basic restrictions in the old theory, the closed-form of the thermodynamic field strength (see ref.\cite{sonnino}), has been removed. Furthermore, the general covariance principle, respected, in reality, only by a very limited class of thermodynamic processes, has been replaced by the thermodynamic covariance principle, first introduced by Prigogine for treating non equilibrium chemical reactions. The validity of the Prigogine statement has been successfully tested, without exception until now, in a wide variety of physical processes going beyond the domain of chemical reactions. The introduction of this principle requested, however, the application of an appropriate mathematical formalism, which may be referred to as the {\it iso-entropic formalism}. The construction of the present theory rests on two assumptions:
\begin{itemize}
\item The thermodynamic theorems valid when a generic thermodynamic system relaxes out of equilibrium are satisfied;
\item There exists a thermodynamic action, scalar under thermodynamic coordinate transformations, which is stationary for general variations in the transport coefficients and the affine connection.
\end{itemize}
\noindent A non-Riemannian geometry has been constructed out of the components of the affine connection, which has been determined by imposing the validity of the Universal Criterion of Evolution for non-equilibrium system relaxing towards a steady-state. Relaxation expresses an intrinsic physical property of a thermodynamic system. The affine connection, on the other hand, is an intrinsic property of geometry allowing to determine the equation for the shortest path. It is the author's opinion that a correct thermodynamical-geometrical theory should correlate these two properties. It is important to mention that the geometry of the thermodynamic space tends to be Riemannian for small values of the inverse of the entropy production. In this limit, we obtain again the same thermodynamic field equations found in ref.\cite{sonnino}. The results established in refs \cite{sonnino3}, for magnetically confined plasmas, and in refs \cite{sonnino4}, for the nonlinear thermoelectric effect and the unimolecular triangular reaction, remain then valid. 

Finally, note that the transport equations may take even more general forms than Eq.~(\ref{ief0}). The fluxes and the forces can be defined locally as fields depending on space coordinates and time. The most general transport relation takes the form
\begin{equation}\label{c1}
J_{\mu}({\bf r},t)=\int_\Omega d{\bf r'}\int_0^t dt' {\mathcal L}_{\mu\nu}(X({\bf r'},t'))X^\nu({\bf r}-{\bf r'},t-t')
\end{equation}
\noindent This type of nonlocal and non Markovian equation expresses the fact that the flux at a given point (${\bf r},t$) could be influenced by the values of the forces in its spatial environment and by its history. Whenever the spatial and temporal ranges of influence are sufficiently small, the delocalization and the retardation of the forces can be neglected under the integral 
\begin{equation}\label{c1a}
{\mathcal L}_{\mu\nu}(X({\bf r'},t'))X^\nu({\bf r}-{\bf r'},t-t')\simeq 
{\mathcal L}_{\mu\nu}(X({\bf r},t))X^\nu({\bf r},t)\delta({\bf r}-{\bf r'})\delta (t-t')
\end{equation}
\noindent and the transport equations reduces to
\begin{equation}\label{c2}
J_{\mu}({\bf r},t)\simeq\tau_{\mu\nu}(X({\bf r},t))X^\nu({\bf r},t)
\end{equation}
\noindent where 
\begin{equation}\label{c3}
\tau_{\mu\nu}(X({\bf r},t))=\int_\Omega d{\bf r'}\int_0^t dt'  {\mathcal L}_{\mu\nu}(X({\bf r},t))
\delta({\bf r}-{\bf r'})\delta (t-t')
\end{equation}
\noindent In the vast majority of cases studied at present in transport theory, it is assumed that the transport equations are of the form of Eq.~(\ref{c2}). However, equations of the form Eq.~(\ref{c1}) may be met when we deal with anomalous transport processes such as, for example, transport in turbulent tokamak plasmas \cite{balescu3}. Eq.~(\ref{c1a}) establishes, in some sort, the limit of validity of the present approach: {\it Eqs~(\ref{tfe1}) determine the nonlinear corrections to the linear ("Onsager") transport coefficients whenever the width of the nonlocal coefficients can be neglected}. Moreover, it is implicitly understood that the thermodynamic quantities (temperature, pressure etc.) are evaluated making use of the local equilibrium principle. This last limitation, however, may be overcame by combining the TFT with the Extended Irreversible Thermodynamics (EIT) \cite{jou}. This will be addressed in the future.

\vskip 0.2truecm
\section{Acknowledgments}
I would like to pay tribute to the memory of Prof. I. Prigogine who gave me the opportunity to exchange most interesting views in different areas  of thermodynamics of irreversible processes. My strong interest in this domain of research is due to him, who promoted the Brussels School of Thermodynamics at the U.L.B., where I took my doctorate in Physics. I am also very grateful to Prof. M. Malek Mansour, from the Universit{\'e} Libre de Bruxelles, Prof. C.M. Becchi and Prof. E. Massa, from the University of Genoa, Dr. F. Zonca, from the EURATOM/ENEA Italian Fusion Association in Frascati (Rome) and Dr. J. Evslin from the SISSA (International School for Advanced Studies) for the useful discussions and suggestions. I would like to thank my hierarchy at the European Commission and the members of the Association-Belgian State for Controlled Thermonuclear Fusion at the U.L.B.
\vskip 0.2truecm
\section{Appendix 1: Transformation Law and Properties of the Affine Connection Eq.(\ref{ts25}).}\label{appx}
\vskip 0.2truecm
In this section we show that the affine connection Eq.~(\ref{ts25}) transforms, under TCT, as in Eq.~(\ref{ts1}) and satisfies the postulates ${\bf 1.}$, ${\bf 2.}$ and ${\bf 3.}$ We first note that the quantity $\delta_\alpha^\lambda\psi_{\beta}+\delta_\beta^\lambda\psi_{\alpha}$ transforms like a mixed thermodynamic tensor of third rank
\begin{equation}\label{appx0}
\delta_\alpha^\lambda\psi_{\beta}+\delta_\beta^\lambda\psi_{\alpha}=(
\delta_\rho^\tau\psi_{\nu}+\delta_\nu^\tau\psi_{\rho})
\frac{\partial X'^\lambda}{\partial X^\tau}\frac{\partial X^\rho}{\partial X'^\alpha}\frac{\partial X^\nu}{\partial X'^\beta}
\end{equation}
\noindent Thus, if Eq.~(\ref{ts9}) transforms, under TCT, like Eq.~(\ref{ts1}), then so will be Eq.~(\ref{ts25}). Consider the symmetric processes. From Eq.~(\ref{tr4}), we have
\begin{equation}\label{appx1}
\!\!\frac{\partial g'_{\alpha\beta}}{\partial X'^\kappa}\!=\!\frac{\partial g_{\rho\nu}}{\partial X^\varrho}\frac{\partial X^\varrho}{\partial X'^\kappa}\frac{\partial X^\rho}{\partial X'^\alpha}\frac{\partial X^\nu}{\partial X'^\beta}+g_{\rho\nu}\frac{\partial^2X^\rho}{\partial X'^\kappa\partial X'^\alpha}\frac{\partial X^\nu}{\partial X'^\beta}+g_{\rho\nu}\frac{\partial^2X^\rho}{\partial X'^\kappa\partial X'^\beta}\frac{\partial X^\nu}{\partial X'^\alpha}\\
\end{equation}
\noindent The thermodynamic Christoffel symbols transform then as 
\begin{equation}\label{appx3}
\begin{Bmatrix} 
\lambda \\ \alpha\beta
\end{Bmatrix}'=
\begin{Bmatrix} 
\tau \\ \rho\nu
\end{Bmatrix}\frac{\partial X'^\lambda}{\partial X^\tau}\frac{\partial X^\rho}{\partial X'^\alpha}\frac{\partial X^\nu}{\partial X'^\beta}+\frac{\partial X'^\lambda}{\partial X^\rho}\frac{\partial^2X^\rho}{\partial X'^\alpha\partial X'^\beta}
\end{equation}
\noindent Recalling that $\sigma'=\sigma$, from Eq.~(\ref{appx1}) we also find
\begin{eqnarray}\label{appx4}
&&\frac{{\bar N}^{'\lambda\kappa}}{2\sigma'}X'_\kappa\mathcal{O}'(g'_{\alpha\beta})=\frac{{\bar N}^{\tau\eta}}{2\sigma}X_\eta\mathcal{O}(g_{\rho\nu})\frac{\partial X'^\lambda}{\partial X^\tau}\frac{\partial X^\rho}{\partial X'^\alpha}\frac{\partial X^\nu}{\partial X'^\beta}\\
&&\frac{1}{2\sigma'}X'^\lambda\mathcal{O}'(g'_{\alpha\beta})=\frac{1}{2\sigma}X^\tau\mathcal{O}(g_{\rho\nu})\frac{\partial X'^\lambda}{\partial X^\tau}\frac{\partial X^\rho}{\partial X'^\alpha}\frac{\partial X^\nu}{\partial X'^\beta}\nonumber
\end{eqnarray}
\noindent where Eqs~(\ref{tr2}) and Eqs~(\ref{tct2}) have been taken into account. Therefore, the affine connection 
\begin{equation}\label{appx5}
{\Gamma}_{\rho\nu}^\tau=
\begin{Bmatrix} 
\tau \\ \rho\nu
\end{Bmatrix}+
\frac{1}{2\sigma}X^\tau\mathcal{O}(g_{\rho\nu})-\frac{1}{2(n+1)\sigma}[\delta_\rho^\tau X^\eta\mathcal{O}(g_{\nu\eta})+\delta_\nu^\tau X^\eta\mathcal{O}(g_{\rho\nu})]
\end{equation}
\noindent transforms as 
\begin{equation}\label{appx6}
{\Gamma'}_{\alpha\beta}^\lambda=
\Gamma_{\rho\nu}^\tau
\frac{\partial X'^\lambda}{\partial X^\tau}\frac{\partial X^\rho}{\partial X'^\alpha}\frac{\partial X^\nu}{\partial X'^\beta}+\frac{\partial X'^\lambda}{\partial X^\rho}\frac{\partial^2X^\rho}{\partial X'^\alpha\partial X'^\beta}
\end{equation}
Consider the general case. From Eq.~(\ref{tr4}) we obtain
\begin{eqnarray}\label{appx7}
\frac{1}{2}\Bigl(\frac{\partial g'_{\alpha\kappa}}{\partial X'^\beta}+\frac{\partial g'_{\beta\kappa}}{\partial X'^\beta}-\frac{\partial g'_{\alpha\beta}}{\partial X'^\kappa}\Bigr)=\!\!\!\!\!\!\!\!&&\frac{\partial X^\varrho}{\partial X'^\kappa}\frac{\partial X^\rho}{\partial X'^\alpha}\frac{\partial X^\nu}{\partial X'^\beta}
\Bigl[\frac{1}{2}\Bigl(\frac{\partial g_{\nu\varrho}}{\partial X^\rho}+\frac{\partial g_{\rho\varrho}}{\partial X^\nu}-\frac{\partial g_{\rho\nu}}{\partial X^\varrho}\Bigr)\Bigr] \nonumber\\
\!\!\!\!\!\!\!\!&&+g_{\rho\nu}\frac{\partial^2 X^\rho}{\partial X'^\alpha\partial X'^\beta}\frac{\partial X^\nu}{\partial X'^\kappa}
\end{eqnarray}
\noindent From Eq.~(\ref{tr6}), we also have
\begin{eqnarray}\label{appx8}
&& \!\!\!\!\frac{\partial f'_{\alpha\mu}}{\partial X'^\beta}\!=\!\frac{\partial f_{\rho\eta}}{\partial X^\varsigma}\frac{\partial X^\varsigma}{\partial X'^\beta}\frac{\partial X^\rho}{\partial X'^\alpha}\frac{\partial X^\eta}{\partial X'^\mu}\!+\! f_{\rho\eta}\frac{\partial^2X^\rho}{\partial X'^\beta\partial X'^\alpha}\frac{\partial X^\eta}{\partial X'^\mu}\!+\!f_{\rho\eta}\frac{\partial^2X^\rho}{\partial X'^\beta\partial X'^\mu}\frac{\partial X^\eta}{\partial X'^\alpha}
\nonumber\\
&&\!\!\!\!\frac{\partial f'_{\beta\mu}}{\partial X'^\alpha}\!=\!\frac{\partial f_{\varsigma\eta}}{\partial X^\rho}\frac{\partial X^\varsigma}{\partial X'^\beta}\frac{\partial X^\rho}{\partial X'^\alpha}\frac{\partial X^\eta}{\partial X'^\mu}\!+\! f_{\rho\eta}\frac{\partial^2X^\rho}{\partial X'^\alpha\partial X'^\beta}\frac{\partial X^\eta}{\partial X'^\mu}\!+\!f_{\rho\eta}\frac{\partial^2X^\rho}{\partial X'^\alpha\partial X'^\mu}\frac{\partial X^\eta}{\partial X'^\beta}\nonumber\\
\end{eqnarray}
\noindent Taking into account Eqs~(\ref{tr2}) and Eqs~(\ref{tct2}) we find
\begin{eqnarray}\label{appx9}
\!\!\!\!\!\!\!\!\!\!\!\!\!\!\!\!&& X'_\kappa X'^\mu\frac{\partial f'_{\alpha\mu}}{\partial X'^\beta}=X_\varrho X^\eta\frac{\partial f_{\rho\eta}}{\partial X^\nu}\frac{\partial X^\varrho}{\partial X'^\kappa}\frac{\partial X^\rho}{\partial X'^\alpha}\frac{\partial X^\nu}{\partial X'^\beta}+X_\nu X^\eta f_{\rho\eta}\frac{\partial^2 X^\rho}{\partial X'^\alpha\partial X'^\beta}\frac{\partial X^\nu}{\partial X'^\kappa} \nonumber\\
\!\!\!\!\!\!\!\!\!\!\!\!\!\!\!\!&&X'_\kappa X'^\mu\frac{\partial f'_{\beta\mu}}{\partial X'^\alpha}=X_\varrho X^\eta\frac{\partial f_{\nu\eta}}{\partial X^\rho}\frac{\partial X^\varrho}{\partial X'^\kappa}\frac{\partial X^\rho}{\partial X'^\alpha}\frac{\partial X^\nu}{\partial X'^\beta}+X_\nu X^\eta f_{\rho\eta}\frac{\partial^2 X^\rho}{\partial X'^\alpha\partial X'^\beta}\frac{\partial X^\nu}{\partial X'^\kappa}\nonumber\\
\end{eqnarray}
\noindent from which we obtain
\begin{eqnarray}\label{appx10}
\frac{1}{2\sigma'}X'_\kappa X'^\mu\Bigl(\frac{\partial f'_{\alpha\mu}}{\partial X'^\beta}+\frac{\partial f'_{\beta\mu}}{\partial X'^\alpha}\Bigr)=\!\!\!\!\!\!\!\!&&\frac{\partial X^\varrho}{\partial X'^\kappa}\frac{\partial X^\rho}{\partial X'^\alpha}\frac{\partial X^\nu}{\partial X'^\beta}\Bigl[\frac{1}{2\sigma}X_\varrho X^\eta\Bigl(\frac{\partial f_{\rho\eta}}{\partial X^\nu}+\frac{\partial f_{\nu\eta}}{\partial X^\rho}\Bigr)\Bigr]\nonumber\\
&& +\frac{1}{\sigma}X_\nu X^\eta f_{\rho\eta}\frac{\partial^2 X^\rho}{\partial X'^\alpha\partial X'^\beta}\frac{\partial X^\nu}{\partial X'^\kappa}
\end{eqnarray}
\noindent where, as usual, $\sigma$ denotes the entropy production. Let us now re-consider the transformations of the following quantities
\begin{eqnarray}\label{appx11}
&& \!\!\!\!\frac{\partial g'_{\alpha\mu}}{\partial X'^\beta}\!=\!\frac{\partial g_{\rho\eta}}{\partial X^\varsigma}\frac{\partial X^\varsigma}{\partial X'^\beta}\frac{\partial X^\rho}{\partial X'^\alpha}\frac{\partial X^\eta}{\partial X'^\mu}\!+\! g_{\rho\eta}\frac{\partial^2X^\rho}{\partial X'^\beta\partial X'^\alpha}\frac{\partial X^\eta}{\partial X'^\mu}\!+\!g_{\rho\eta}\frac{\partial^2X^\rho}{\partial X'^\beta\partial X'^\mu}\frac{\partial X^\eta}{\partial X'^\alpha}
\nonumber\\
&&\!\!\!\!\frac{\partial g'_{\beta\mu}}{\partial X'^\alpha}\!=\!\frac{\partial g_{\varsigma\eta}}{\partial X^\rho}\frac{\partial X^\varsigma}{\partial X'^\beta}\frac{\partial X^\rho}{\partial X'^\alpha}\frac{\partial X^\eta}{\partial X'^\mu}\!+\! g_{\rho\eta}\frac{\partial^2X^\rho}{\partial X'^\alpha\partial X'^\beta}\frac{\partial X^\eta}{\partial X'^\mu}\!+\!g_{\rho\eta}\frac{\partial^2X^\rho}{\partial X'^\alpha\partial X'^\mu}\frac{\partial X^\eta}{\partial X'^\beta}\nonumber\\
\end{eqnarray}
\noindent From these equations we obtain
\begin{equation}\label{appx12}
X'^\mu\frac{\partial g'_{\alpha\mu}}{\partial X'^\beta}+X'^\mu\frac{\partial g'_{\beta\mu}}{\partial X'^\alpha}\!=\!\Bigl(X^\eta\frac{\partial g_{\rho\eta}}{\partial X^\varsigma}+
\frac{\partial g_{\varsigma\eta}}{\partial X^\rho}\Bigr)\frac{\partial X^\varsigma}{\partial X'^\beta}\frac{\partial X^\rho}{\partial X'^\alpha}\frac{\partial X^\eta}{\partial X'^\mu}\!+\! 2X_\rho\frac{\partial^2X^\rho}{\partial X'^\beta\partial X'^\alpha}
\end{equation}
\noindent where Eqs~(\ref{tct2}) have been taken into account. From Eq.~(\ref{appx12}) we finally obtain
\begin{eqnarray}\label{appx13}
&&\!\!\!\!\!\!\!\!\!\!\!\!\!\!\!\!\!\!\!\!\!\!\!\!\!\!\!\!\frac{1}{2\sigma'}\Bigl[X'^\varsigma\Bigl(\frac{\partial g'_{\alpha\varsigma}}{\partial X'^\beta}+\frac{\partial g'_{\beta\varsigma}}{\partial X'^\alpha}\Bigr)\Bigr]f'_{\kappa\mu}X'^\mu=\nonumber\\
&&\qquad\qquad\quad\frac{1}{2\sigma}\Bigl[X^\eta\Bigl(\frac{\partial g_{\rho\eta}}{\partial X^\nu}+\frac{\partial g_{\nu\eta}}{\partial X^\rho}\Bigr)\Bigr]f_{\varrho\varsigma}X^\varsigma\frac{\partial X^\varrho}{\partial X'^\kappa}\frac{\partial X^\rho}{\partial X'^\alpha}\frac{\partial X^\nu}{\partial X'^\beta}
\nonumber\\
&&\qquad\qquad\quad+\frac{1}{\sigma}X_\rho X^\eta f_{\nu\eta}\frac{\partial^2 X^\rho}{\partial X'^\alpha\partial X'^\beta}\frac{\partial X^\nu}{\partial X'^\kappa}
\end{eqnarray}
\noindent Summing Eq.~(\ref{appx7}) with Eqs~(\ref{appx10}) and (\ref{appx13}), it follows that
\begin{equation}\label{appx14}
{\tilde\Gamma}_{\alpha\beta}^{\prime\lambda}=
{\tilde \Gamma}_{\rho\nu}^\tau
\frac{\partial X'^\lambda}{\partial X^\tau}\frac{\partial X^\rho}{\partial X'^\alpha}\frac{\partial X^\nu}{\partial X'^\beta}+\frac{\partial X'^\lambda}{\partial X^\rho}\frac{\partial^2 X^\rho}{\partial X'^\alpha\partial X'^\beta}
\end{equation}
\noindent where 
\begin{eqnarray}\label{appx15}
\!\!\!\!\!\!\!
{\tilde \Gamma}_{\rho\nu}^\tau=\!\!\!\!\!\!\!\!\!\!&&{\bar N}^{\tau\varrho}g_{\varrho\varsigma}\begin{Bmatrix} 
\varsigma \\ \rho\nu
\end{Bmatrix}+\frac{{\bar N}^{\tau\varrho} X_{\varrho}X^\eta}{2\sigma}\Bigl(\frac{\partial f_{\rho\eta}}{\partial X^\nu}+\frac{\partial f_{\nu\eta}}{\partial X^\rho}\Bigr)
\nonumber\\
&&\qquad\qquad\qquad\qquad\qquad\qquad
+\frac{{\bar N}^{\tau\varrho}f_{\varrho\varsigma}X^\varsigma X^\eta}{2\sigma}
\Bigr(\frac{\partial g_{\rho\eta}}{\partial X^\nu}+\frac{\partial g_{\nu\eta}}{\partial X^\rho}\Bigl)
\end{eqnarray}
\noindent and 
\begin{equation}\label{appx16}
{\bar N}^{\tau\varrho}N_{\rho\varrho}=\delta_{\rho}^{\tau}\qquad {\rm}\quad {\rm with}\quad N_{\rho\varrho}=g_{\rho\varrho}+\frac{1}{\sigma}f_{\rho\eta}X^\eta X_\varrho+\frac{1}{\sigma}f_{\varrho\eta}X^\eta X_\rho
\end{equation}
\noindent Summing again Eq.~(\ref{appx15}) with Eq.~(\ref{appx0}) and the first equation of Eq.~(\ref{appx4}), we finally obtain
\begin{equation}\label{appx17}
{\Gamma'}_{\alpha\beta}^\lambda=
\Gamma_{\rho\nu}^\tau
\frac{\partial X'^\lambda}{\partial X^\tau}\frac{\partial X^\rho}{\partial X'^\alpha}\frac{\partial X^\nu}{\partial X'^\beta}+\frac{\partial X'^\lambda}{\partial X^\rho}\frac{\partial^2X^\rho}{\partial X'^\alpha\partial X'^\beta}
\end{equation}
\noindent where
\begin{equation}\label{appx18}
\Gamma_{\rho\nu}^\tau={\tilde \Gamma}_{\rho\nu}^\tau+\frac{{\bar N}^{\tau\eta}}{2\sigma}X_\eta\mathcal{O}(g_{\rho\nu})+\delta_\rho^\tau\psi_{\nu}+\delta_\nu^\tau\psi_{\rho}
\end{equation}
\noindent It is not difficult to prove that the affine connection Eq.~(\ref{ts25}) satisfies the postulates ${\bf 1.}$, ${\bf 2.}$ and ${\bf 3.}$ Indeed, if $A^\mu$ indicates a thermodynamic vector, we have 
\begin{equation}\label{appx19}
A'^\lambda=A^\eta\frac{\partial X'^\lambda}{\partial X^\eta}
\end{equation}
\noindent Deriving this equation, with respect to parameter $\varsigma$, we obtain
\begin{equation}\label{appx20}
\frac{dA'^\lambda}{d\varsigma}=\frac{dA^\eta}{d\varsigma}\frac{\partial X'^\lambda}{\partial X^\eta}+A^\eta\frac{\partial^2 X'^\lambda}{\partial X^\tau\partial X^\eta}\frac{dX^\tau}{d\varsigma}
\end{equation}
\noindent Taking into account the following identities 
\begin{equation}\label{appx20a}
\frac{\partial^2 X'^\lambda}{\partial X^\tau\partial X^\eta}=-\frac{\partial X'^\lambda}{\partial X^\rho}\frac{\partial X'^\alpha}{\partial X^\tau}\frac{\partial^2 X^\rho}{\partial X^\eta\partial X'^\alpha}=-\frac{\partial X'^\alpha}{\partial X^\tau}\frac{\partial X'^\beta}{\partial X^\eta}\frac{\partial X'^\lambda}{\partial X^\rho}\frac{\partial^2 X^\rho}{\partial X'^\alpha\partial X'^\beta}
\end{equation}
\noindent and Eq.~(\ref{appx17}), we find
\begin{equation}\label{appx21}
\frac{\delta A'^\lambda}{\delta \varsigma}=\frac{\delta A^\eta}{\delta \varsigma}\frac{\partial X'^\lambda}{\partial X^\eta}
\end{equation}
\noindent The validity of postulates ${\bf 2.}$ and ${\bf 3.}$ is immediately verified, by direct computation, using Eqs~(\ref{ts2}) and (\ref{ts5}). The validity of these postulates was shown above for a thermodynamic vector. By a closely analogous procedure it can be checked that the postulated ${\bf 1.}$ , ${\bf 2.}$ and ${\bf 3.}$ are satisfied for any thermodynamic tensor.
\vskip 0.2truecm
\section{Appendix 2: Derivation of the Thermodynamic Field Equations from the Action Principle.}\label{appeq}
\vskip 0.2truecm
In this appendix, the thermodynamic field equations by the principle of the least action are derived. Let us rewrite Eq.~(\ref{pa5}) as
\begin{equation}\label{appeq1}
I=\int\Bigl[ R_{\mu\nu}g^{\mu\nu}-(\Gamma^\lambda_{\mu\nu}-
{\tilde\Gamma}^\lambda_{\mu\nu})S^{\mu\nu}_\lambda
\Bigr]\sqrt{g}\ \! {d^{}}^n\!X
\end{equation}
\noindent where the expression of $S^{\mu\nu}_\lambda$ is given by Eq.~(\ref{pa4}). This action is stationary by varying independently the transport coefficients (i.e. by varying, separately, $g_{\mu\nu}$ and $f_{\mu\nu}$) and the affine connection $\Gamma^\lambda_{\mu\nu}$. A variation with respect to $\Gamma^\lambda_{\mu\nu}$ reads
\begin{equation}\label{appeq2}
\delta I_{\Gamma}=\int\Bigl[ \delta R_{\mu\nu}g^{\mu\nu}-\delta\Gamma^\lambda_{\mu\nu}S^{\mu\nu}_\lambda
\Bigr]\sqrt{g}\ \! {d^{}}^n\!X=0
\end{equation}
\noindent By direct computation, we can check that
\begin{equation}\label{appeq3}
\delta R_{\mu\nu}=(\delta\Gamma^\lambda_{\mu\lambda})_{\mid\nu}-(\delta\Gamma^\lambda_{\mu\nu})_{\mid\lambda}
\end{equation}
\noindent Defining $\mathcal{K}^{\mu\nu}\equiv\sqrt{g}g^{\mu\nu}$, we have the identities
\begin{eqnarray}\label{appeq4}
&&(\mathcal{K}^{\mu\nu}\delta\Gamma^\lambda_{\mu\lambda})_{\mid\nu}=\mathcal{K}^{\mu\nu}_{\mid\nu}\delta\Gamma^\lambda_{\mu\lambda}+\mathcal{K}^{\mu\nu}\delta\Gamma^\lambda_{\mu\lambda\mid\nu}\nonumber\\
&&
(\mathcal{K}^{\mu\nu}\delta\Gamma^\lambda_{\mu\nu})_{\mid\lambda}=\mathcal{K}^{\mu\nu}_{\mid\lambda}\delta\Gamma^\lambda_{\mu\nu}+\mathcal{K}^{\mu\nu}\delta\Gamma¬\lambda_{\mu\nu\mid\lambda}
\end{eqnarray}
\noindent Eq.~(\ref{appeq2}) can be rewritten as
\begin{eqnarray}\label{appeq5}
\delta I_{\Gamma}=&&\!\!\!\!\!\!\!\!\!\!\!\!
\int(\mathcal{K}^{\mu\nu}\delta\Gamma^\lambda_{\mu\lambda})_{\mid\nu}{d^{}}^n\!X-\int\mathcal{K}^{\mu\nu}_{\mid\nu}\delta\Gamma^\lambda_{\mu\lambda}{d^{}}^n\!X+\int
\mathcal{K}^{\mu\nu}_{\mid\lambda}\delta\Gamma^\lambda_{\mu\nu}{d^{}}^n\!X-
\nonumber\\
&&\!\!\!\!\!\!\!\!\!\!\!\!
\int(\mathcal{K}^{\mu\nu}\delta\Gamma^\lambda_{\mu\nu})_{\mid\lambda}{d^{}}^n\!X
-\int S^{\mu\nu}_\lambda\delta\Gamma^\lambda_{\mu\nu}
\sqrt{g}\ \! {d^{}}^n\!X=0
\end{eqnarray}
\noindent The thermodynamic covariant derivative of the metric tensor reads
\begin{equation}\label{appeq14}
g_{\alpha\beta\mid\lambda}=g_{\alpha\beta,\lambda}-\Gamma^\eta_{\alpha\lambda}g_{\eta\beta}-\Gamma^\eta_{\beta\lambda}g_{\eta\alpha}
\end{equation}
\noindent from which we find
\begin{equation}\label{appeq15}
\Gamma^\beta_{\lambda\beta}=-\frac{1}{2}g^{\alpha\beta}g_{\alpha\beta\mid\lambda}+\frac{1}{2}g^{\alpha\beta}g_{\alpha\beta,\lambda}
\end{equation}
\noindent Taking into account that $\delta\sqrt{g}=1/2\sqrt{g}g^{\mu\nu}\delta g_{\mu\nu}$, Eq.~(\ref{appeq15}) can also be brought into the form 
\begin{equation}\label{appeq15a}
\Gamma^\beta_{\lambda\beta}-\frac{1}{\sqrt{g}}\sqrt{g}_{,\lambda}+\frac{1}{\sqrt{g}}\sqrt{g}_{\mid\lambda}=0
\end{equation}
\noindent On the other hand, we can easily check the validity of the following identities
\begin{eqnarray}\label{appeq7}
&&\!\!\!\!\!\!\!\!\!\!\!\!\!\!\!\!
(\mathcal{K}^{\mu\nu}\delta\Gamma^\lambda_{\mu\lambda})_{\mid\nu}=(\mathcal{K}^{\mu\nu}\delta\Gamma^\lambda_{\mu\lambda})_{,\nu}
+(\Gamma^\beta_{\nu\beta}-\frac{1}{\sqrt{g}}\sqrt{g}_{,\nu}+\frac{1}{\sqrt{g}}\sqrt{g}_{\mid\nu}){\mathcal{K}}^{\mu\nu}\delta\Gamma^\lambda_{\mu\lambda}
\nonumber\\
&&\!\!\!\!\!\!\!\!\!\!\!\!\!\!\!\!
(\mathcal{K}^{\mu\nu}\delta\Gamma^\lambda_{\mu\nu})_{\mid\lambda}=(\mathcal{K}^{\mu\nu}\delta\Gamma^\lambda_{\mu\nu})_{,\lambda}
+(\Gamma^\beta_{\lambda\beta}-\frac{1}{\sqrt{g}}\sqrt{g}_{,\lambda}+\frac{1}{\sqrt{g}}\sqrt{g}_{\mid\lambda}){\mathcal{K}}^{\mu\nu}\delta\Gamma^\lambda_{\mu\nu}
\end{eqnarray}
\noindent Therefore, from Eq.~(\ref{appeq15a}), the terms
\begin{equation}\label{appeq6}
\int(\mathcal{K}^{\mu\nu}\delta\Gamma^\lambda_{\mu\lambda})_{\mid\nu}{d^{}}^n\!X
\qquad{\rm and}\qquad \int(\mathcal{K}^{\mu\nu}\delta\Gamma^\lambda_{\mu\nu})_{\mid\lambda}{d^{}}^n\!X
\end{equation}
\noindent drop out when we integrate over all thermodynamic space. Eq.~(\ref{appeq5}) reduces then to
\begin{equation}\label{appeq9}
\delta I_{\Gamma}=
-\int\mathcal{K}^{\mu\nu}_{\mid\nu}\delta\Gamma^\lambda_{\mu\lambda}{d^{}}^n\!X+\int
\mathcal{K}^{\mu\nu}_{\mid\lambda}\delta\Gamma^\lambda_{\mu\nu}{d^{}}^n\!X-\int S^{\mu\nu}_\lambda\delta\Gamma^\lambda_{\mu\nu}
\sqrt{g}\ \! {d^{}}^n\!X=0
\end{equation}
\noindent It is seen that $\delta I_\Gamma$ vanishes for general variation of $\delta\Gamma^\lambda_{\mu\nu}$ if, and only if, 
\begin{equation}\label{appeq10}
-\frac{1}{2}\mathcal{K}^{\mu\alpha}_{\mid\alpha}\delta^\nu_\lambda
-\frac{1}{2}\mathcal{K}^{\nu\alpha}_{\mid\alpha}\delta^\mu_\lambda+
\mathcal{K}^{\mu\nu}_{\mid\lambda}-
S^{\mu\nu}_\lambda\sqrt{g}=0
\end{equation}
\noindent Contracting indexes $\nu$ with $\lambda$, we find
\begin{equation}\label{appeq11}
\mathcal{K}^{\mu\alpha}_{\mid\alpha}-\Psi^\mu_{\alpha\beta}g^{\alpha\beta}\sqrt{g}=0
\end{equation}
\noindent where Eq.~(\ref{pa4}) has been taken into account. Thanks to Eq.~(\ref{appeq11}), Eq.~(\ref{appeq10}) becomes
\begin{equation}\label{appeq12}
\mathcal{K}^{\mu\nu}_{\mid\lambda}=\Psi^\mu_{\alpha\lambda}g^{\nu\alpha}\sqrt{g}+\Psi^\nu_{\alpha\lambda}g^{\mu\alpha}\sqrt{g}
\end{equation}
\noindent From the identity $\delta g^{\mu\nu}=-g^{\mu\alpha}g^{\nu\beta}\delta g_{\alpha\beta}$, we also have 
\begin{equation}\label{appeq13}
\mathcal{K}^{\mu\nu}_{\mid\lambda}=\sqrt{g}_{\mid\lambda}g^{\mu\nu}+\sqrt{g}g^{\mu\nu}_{\mid\lambda}=\frac{1}{2}\sqrt{g}g^{\mu\nu}g^{\alpha\beta}g_{\alpha\beta\mid\lambda}-\sqrt{g}g^{\mu\alpha}g^{\nu\beta}g_{\alpha\beta\mid\lambda}
\end{equation}
\noindent Eq.~(\ref{appeq12}) reads then
\begin{equation}\label{appeq16}
-g^{\mu\alpha}g^{\nu\beta}g_{\alpha\beta\mid\lambda}+\frac{1}{2}g^{\alpha\beta}g_{\alpha\beta\mid\lambda}g^{\mu\nu}=
\Psi^\mu_{\alpha\lambda}g^{\nu\alpha}+\Psi^\nu_{\alpha\lambda}g^{\mu\alpha}
\end{equation}
\noindent Contracting this equation with $g_{\mu\nu}$, we find, for $n\neq2$
\begin{equation}\label{appeq17}
g^{\alpha\beta}g_{\alpha\beta\mid\lambda}=0
\end{equation}
\noindent where Eqs~(\ref{pa2}) have been taken into account. Eq.~(\ref{appeq16}) is simplified as
\begin{equation}\label{appeq18}
-g^{\mu\alpha}g^{\nu\beta}g_{\alpha\beta\mid\lambda}=
\Psi^\mu_{\alpha\lambda}g^{\nu\alpha}+\Psi^\nu_{\alpha\lambda}g^{\mu\alpha}
\end{equation}
\noindent Contracting again Eq.~(\ref{appeq18}) with $g_{\mu\eta}g_{\nu\rho}$, we finally obtain
\begin{equation}\label{appeq19}
g_{\eta\rho\mid\lambda}=-\Psi^\alpha_{\eta\lambda}g_{\alpha\rho}-\Psi^\alpha_{\rho\lambda}g_{\alpha\eta}
\end{equation}
\noindent The first two field equations in Eqs~(\ref{tfe1}) are straightforwardly obtained considering that from Eq.~(\ref{appeq19}) we derive $\Gamma^\lambda_{\mu\nu}-{\tilde \Gamma}^\lambda_{\mu\nu}=0$ (see section \ref{isoef}).



\end{document}